\documentclass[11pt]{JHEP3}
\usepackage{latexsym}
\usepackage{epsfig}
\usepackage{amssymb}
\usepackage{amsthm}
\usepackage{amsmath}
\usepackage{amssymb,amsfonts}
\usepackage{verbatim}

\DeclareMathOperator{\tr}{\textrm{tr}}
\DeclareMathOperator{\Tr}{\textrm{Tr}}
\DeclareMathOperator{\Str}{\textrm{Str}}
\DeclareMathOperator{\diag}{\textrm{diag}}

\DeclareMathOperator{\MA}{\mathcal{A}}

\DeclareMathOperator{\MF}{\mathcal{F}}
\DeclareMathOperator{\MG}{\mathcal{G}}

\DeclareMathOperator{\MM}{\mathcal{M}}
\DeclareMathOperator{\MN}{\mathcal{N}}

\DeclareMathOperator{\MR}{\mathcal{R}}

\DeclareMathOperator{\MBFA}{\mathbf{A}}
\DeclareMathOperator{\MBFC}{\mathbf{C}}
\DeclareMathOperator{\MBFG}{\mathbf{G}}
\DeclareMathOperator{\MBFS}{\mathbf{S}}

\DeclareMathOperator{\mbfb}{\mathbf{b}}
\DeclareMathOperator{\mbfv}{\mathbf{v}}

\DeclareMathOperator{\MBBR}{\mathbb{R}}
\DeclareMathOperator{\MBBZ}{\mathbb{Z}}

\newcommand{\nn}{\nonumber}

\title{A systematic search for anomaly-free supergravities in six dimensions}

\author{Spyros D. Avramis and Alex Kehagias\\
    Department of Physics, National Technical University of Athens, GR-15773, Zografou, Athens,
    Greece\\
    E-mail: \email{avramis@cern.ch}, \email{kehagias@cern.ch}}

\preprint{\hepth{0508172}}

\abstract{We conduct a systematic search for anomaly-free six-dimensional $\MN=1$ chiral supergravity theories. Under a certain set of restrictions on the allowed gauge groups and the representations of the hypermultiplets, we enumerate all possible Poincar\'e and gauged supergravities with one tensor multiplet satisfying the 6D anomaly cancellation criteria.}

\begin{document}

\allowdisplaybreaks{

\section{Introduction}
\label{sec-1}

Anomaly cancellation has been, for a long time, one of the main guiding principles for the construction of consistent gauge and gravitational theories. The most familiar examples of anomaly cancellation in four-dimensional theories are the automatic cancellation of gauge and mixed abelian anomalies in the Standard Model as well as the necessity for the introduction of a second Higgs doublet to achieve the same type of cancellation in the MSSM. However, in these cases, anomaly cancellation cannot by itself provide serious constraints on the gauge group and the particle spectrum of the theory, since the cancellation conditions are weak and can be satisfied by a vast number of models. On the other hand, in the case of theories living in ten or six dimensions where gravitational and mixed nonabelian anomalies \cite{Alvarez-Gaume:1983ig,Alvarez-Gaume:1984dr} are present in addition to gauge and mixed abelian ones, the requirement of anomaly cancellation may lead to powerful constraints singling out a relatively small number of consistent models. In the particularly interesting case of theories considered as long-wavelength limits of fundamental theories whose detailed structure is not fully known, the search for anomaly-free models is greatly motivated by the fact that anomaly cancellation, being an infrared effect, may enable us to infer information about the high-energy aspects of the underlying theory through low-energy considerations.

In this respect, it is instructive to recall the basic facts in the case of 10D supergravity. The string-derived chiral 10D supergravities known before the explicit calculation of higher-dimensional anomalies were Type IIB $\MN=2$ supergravity (realized in terms of closed strings) and $\MN=1$ supergravity coupled to $\textrm{SO}(N)$ Yang-Mills (realized in terms of Type I strings). A striking result of the calculation of 10D gravitational anomalies in \cite{Alvarez-Gaume:1983ig} was the complete cancellation of anomalies for the Type IIB theory; on the other hand, $\MN=1$ supergravity was found to be anomalous. However, Green and Schwarz discovered that $\MN=1$ theory can also be made anomaly-free \cite{Green:1984sg} through a coupling of the 2--form of the supergravity multiplet to a certain 8--form constructed out of curvature invariants. The necessary and sufficient condition for anomaly cancellation was that the anomaly polynomial must factorize. This can happen only for a gauge group of dimension 496 with no sixth-order Casimirs, in which case the factorization coefficients are uniquely determined and result in a further constraint on certain group-theoretical coefficients. The obvious candidate was $\textrm{SO}(32)$ which indeed satisfied all the above requirements and the corresponding string theory was subsequently shown \cite{Green:1984ed} to also satisfy the RR tadpole cancellation condition. However, surprisingly enough, these requirements were also satisfied by the $E_8 \times E_8$ group which at that time lacked a string-theoretical interpretation, as well as by the physically uninteresting $E_8 \times \textrm{U}(1)^{248}$ and $\textrm{U}(1)^{496}$ groups; the above four groups exhaust all possibilities. The discovery of the heterotic string provided a string realization of the $E_8 \times E_8$ model which turned out to be the most phenomenologically relevant string unification model at the time. These developments made clear that anomaly cancellation not only seriously constrains the particle spectrum of a theory but can also point, from the effective-field-theory point of view, to new consistent models that may be realized through a more fundamental theory.

The Green-Schwarz mechanism also carries over to lower-dimensional chiral theories like the minimal 6D supergravities \cite{Nishino:1984gk,Nishino:1986dc,Nishino:1997ff,Ferrara:1997gh,Riccioni:2001bg};
the relevant anomaly cancellation conditions are discussed in \cite{Salam:1985mi, Bergshoeff:1986hv,Ketov:1989an,Ketov:1990jr,Erler:1993zy,Schwarz:1995zw}. In the 6D case, things are more complicated mainly due to the existence of the massless hypermultiplets that may transform in arbitrary representations of the gauge group. A consequence of this is that the anomaly cancellation conditions are somewhat weaker than those in the 10D case. First, the condition for the cancellation of irreducible gravitational anomalies does not uniquely fix the dimension of the gauge group but, instead, it simply sets an upper bound on the number of non-singlet hypermultiplets. Second, in the case that the gauge group has fourth-order Casimirs, cancellation of the corresponding irreducible gauge anomaly leads to an equality constraint for the numbers of hypermultiplets. Finally, the factorization condition does not determine how the highest-order traces in the gauge anomaly must factorize but instead leads to two weaker constraints. The conditions mentioned above admit a large number of solutions for the gauge group and the possible hypermultiplet representations and, in fact, a complete classification is a very complicated task. In the related literature, a relatively small number of the possible theories has been explored.

The search for consistent six-dimensional $\MN=1$ supergravities is greatly motivated by a number of reasons, namely (i) their shared properties with ten-dimensional $\MN=1$ supergravities, (ii) their use as toy models for the study of complicated phenomena such as flux compactifications, (iii) their connection, in the gravity-decoupling limit, to the much-studied $\MN=2$ supersymmetric gauge theories in four dimensions, (iv) the possibility of vectorlike \cite{Salam:1984cj} or chiral \cite{Randjbar-Daemi:1985wc} compactifications of the gauged theories down to flat four-dimensional space using a gauge field residing in an internal $\MBFS^2$ and (v) the partial solution they provide to the cosmological constant problem in both ungauged \cite{Kehagias:2004fb,Nair:2004yu,Randjbar-Daemi:2004ni} and gauged \cite{Aghababaie:2003wz,Aghababaie:2003ar,Burgess:2004dh} cases.

Regarding the case of Poincar\'e (ungauged) supergravities, most models found so far correspond to heterotic string compactifications on $K3$ \cite{Green:1984bx}, possibly involving symmetry enhancement either from the Gepner points of orbifold realizations of $K3$ \cite{Erler:1993zy} or by small instantons \cite{Schwarz:1995zw,Witten:1995gx}, as well as chains of models obtained from the above ones by Higgsing. In \cite{Schwarz:1995zw}, a few more models were found by directly solving the anomaly-cancellation conditions. Finally, many series of models were constructed \cite{Bershadsky:1996nh,Bershadsky:1997sb} using geometric engineering via F-theory. Moreover, the issue of anomaly cancellation in six dimensions has been examined in some slightly different classes of theories. One such class corresponds to boundary theories in Ho\v rava-Witten--type compactifications of 7D supergravity on $\MBFS^1/\MBBZ_2$ \cite {Gherghetta:2002xf,Gherghetta:2002nq,Avramis:2004cn}. Another class corresponds to flat-space 6D gauge theories, where anomaly cancellation is related to the existence of non-trivial RG fixed points \cite{Seiberg:1996qx,Danielsson:1997kt}. Although, the number of known anomaly-free 6D Poincar\'e supergravities is quite large, it is certainly useful to tabulate the simplest of them and it is interesting to search whether there are more anomaly-free models or chains of models than those already found.

Turning to the gauged case, the known anomaly-free models are an $E_7 \times E_6 \times \textrm{U}(1)_R$ model found in \cite{Randjbar-Daemi:1985wc} as well as a recently-discovered $E_7 \times G_2 \times \textrm{U}(1)_R$ model \cite{Avramis:2005qt}. There are also a few models \cite{Salam:1985mi} involving extra ``drone'' $\textrm{U}(1)$'s. These models have been found from purely supergravity considerations, guided by the requirement of anomaly cancellation. The uniqueness of these models and their interesting physical properties provide a great motivation for investigating whether more models of this type exist.

It is the purpose of this paper to partially address the two problems mentioned in the preceding two paragraphs. Here, we present a search for anomaly-free six-dimensional $\MN=1$ supergravities, subject to a certain set of conditions on the allowed gauge groups and representations. Under these conditions, we do an exhaustive search for anomaly-free models, that is, for models where the Green-Schwarz mechanism may operate. In the course of the search, various known models are identified while there appear models not previously found.

The paper is organized as follows. In Section 2 we review the basic facts about anomalies in $D=6$, $\MN=1$ supergravity theories, we state the conditions for cancellation of local anomalies and absence of global gauge anomalies and we state the restrictions for our search. In Section 3, we enumerate the anomaly-free Poincar\'e supergravities found in our search, while in Section 4 we do the same for gauged supergravities. Finally, in Section 5, we discuss our main results.

\section{Review of anomaly cancellation in six dimensions}
\label{sec-2}

In this section, we fix our notation and conventions, we describe the basics of $D=6$, $\MN=1$ supergravities and we give a review of anomaly cancellation in these theories. The anomaly cancellation mechanism is presented in full detail and includes discussions of gauged theories, the generalized Green-Schwarz mechanism and global anomalies. The aim is to provide a self-contained treatment that may facilitate further search for anomaly-free models.

\subsection{Basics of $D=6$, $\MN=1$ supergravity}
\label{sec-2-1}

The minimal $\MN=1$ supersymmetry algebra in six dimensions is chiral and has $\textrm{Sp}(1)$ as its R-symmetry group. Its massless representations, classified in terms of the $\textrm{SO}(4) \cong \textrm{SU}(2) \times \textrm{SU}(2)$ little group and the $\textrm{Sp}(1)$ R-symmetry group, and their particle content are\footnote{Our conventions are appropriate for the signature $(-,+,+,+,+,+)$ for the spacetime metric.}:
\begin{eqnarray}
\label{e-2-1}
\text{Supergravity multiplet}
\quad&:&\quad (\mathbf{3},\mathbf{3};\mathbf{1}) +
(\mathbf{1},\mathbf{3};\mathbf{1}) +
(\mathbf{2},\mathbf{3};\mathbf{2})
= ( g_{\mu\nu} , B^+_{\mu\nu} , \psi^{i-}_\mu ),\nn\\
\text{Tensor multiplet} \quad&:&\quad
(\mathbf{3},\mathbf{1};\mathbf{1}) + (\mathbf{1},\mathbf{1};\mathbf{1}) + (\mathbf{1},\mathbf{2};\mathbf{2})
= ( B^-_{\mu\nu} , \phi , \chi^{i+} ), \nn\\
\text{Vector multiplet} \quad&:&\quad
(\mathbf{2},\mathbf{2};\mathbf{1}) + (\mathbf{2},\mathbf{1};\mathbf{2})
= ( A_\mu, \lambda^{i-} ), \nn\\
\text{Hypermultiplet} \quad&:&\quad 4
(\mathbf{1},\mathbf{1};\mathbf{2}) + 2
(\mathbf{1},\mathbf{2};\mathbf{1}) = ( 4 \varphi , 2 \psi^{+} ).
\end{eqnarray}
Here, the spinors are symplectic Majorana, the index $i=1,2$ takes values in the fundamental of $\textrm{Sp}(1)$ and the $+$ ($-$) superscripts denote positive (negative) chirality for spinors and (anti-)self-duality for 2--forms.

A general $D=6$, $\MN=1$ supergravity theory coupled to matter is constructed by combining one supergravity multiplet with $n_T$ tensor multiplets, $n_V$ vector multiplets and $n_H$ hypermultiplets, where $n_T$, $n_V$ and $n_H$ are defined so as to include group multiplicities. The $n_T$ real scalars in the tensor multiplet parameterize the coset space $\textrm{SO}(1,n_T) / \textrm{SO}(n_T)$. The $4 n_H$ real hyperscalars parameterize a non-compact quaternionic manifold of the form
\begin{equation}
\label{e-2-2} \MM = \frac{G}{H \times \textrm{Sp}(1) },
\end{equation}
where the $\textrm{Sp}(1)$ subgroup is identified with the R-symmetry group, while the hyperinos furnish an appropriate representation of $H$. The allowed choices for $(G,H)$ \cite{Nishino:1986dc} are given by $( \textrm{Sp}(n_H,1) , \textrm{Sp}(n_H) )$, $( \textrm{SU}(n_H,2) , \textrm{SU}(n_H) \times \textrm{U}(1) )$, $(\textrm{SO}(n_H,4) , \textrm{SO}(n_H) \times \textrm{SO}(3) )$, $( E_8 , E_7 )$, $( E_7 , \textrm{SO}(12) )$, $( E_6 , \textrm{SU}(6) )$, $( F_4 , \textrm{Sp}(3) )$ and $( G_2, \textrm{Sp}(1) )$. The vector multiplets may belong to a gauge group $\MG$ which is the product of a subgroup of the isometry group $G$ and a possible ``shadow'' group $S$ under which all other multiplets are inert. In the first three cases, where $G$ is non-compact, this essentially means\footnote{For the remaining cases where $G$ is compact, the gauge group can be any subgroup of $G$ times $S$ but the hyperinos are restricted to transform only under $H$.} that $\MG$ is a subgroup of the $H \times \textrm{Sp}(1)$ holonomy group times $S$. For the analysis that follows, it is convenient to write this gauge group as $\MG = \MG_s \times \MG_r \times \MG_a$, where (i) $\MG_s$ is a semisimple group containing factors from $H$ and $S$ given by the product $\prod_\alpha \MG_\alpha$ where the $\MG_\alpha$'s are simple, (ii) $\MG_r$ is the R-symmetry factor arising in gauged theories and can be either $\textrm{Sp}(1)$ or a $\textrm{U}(1)$ subgroup thereof and (iii) $\MG_a$ is an abelian subgroup of $S$ (abelian factors arising from $H$ would only make sense if resulting from a fundamental model so that the charges would be fixed). Introducing an extended index $A=1,\ldots,N$ that runs over all group factors in $\MG_s \times \MG_r$ (i.e. $A=\alpha$ for ungauged theories and $A =( \alpha,r )$ for gauged theories), we write the full group as $\MG = ( \prod_A \MG_A ) \times \MG_a$.

The transformation properties of the various fermions under the gauge group are as follows. Under $\MG_s$, the hyperinos may transform in arbitrary representations while the gravitino and tensorinos are inert. Under $\MG_r$, the hyperinos are inert (although the hyperscalars are charged) while the gravitino, tensorinos and gauginos transform non-trivially. In particular, in the case where the whole $\textrm{Sp}(1)$ is gauged, Eq. (\ref{e-2-1})
indicates that the gravitino, the tensorinos and the $\MG_s$ gauginos transform in the fundamental $\mathbf{2}$ while the $\textrm{Sp}(1)$ gauginos transform in the $\mathbf{3} \otimes \mathbf{2} = \mathbf{2} \oplus \mathbf{4}$. In the case where only a $\textrm{U}(1) \subset \textrm{Sp}(1)$ is gauged, the gravitino, the tensorinos and all gauginos have unit charge.

Let us now write down the spectrum of the theories under consideration. Starting from the tensor multiplets, we will keep $n_T$ arbitrary, bearing in mind that the case $n_T=1$ is rather special in the respect that it is the only one for which a covariant Lagrangian formulation of the theory is possible and the only one that can result from the usual perturbative heterotic string compactifications. As for the vector multiplets, their total number is given by
\begin{equation}
\label{e-2-3}
n_V = \dim \MG = \dim \MG_s + \dim \MG_r + \dim \MG_a.
\end{equation}
Finally, for the hypermultiplets, we denote a generic representation of $\MG_s$ by $(\MR_{i_1}, \ldots , \MR_{i_k})$ and we let $n_{i_1 \ldots i_k}$ be the number of hypermultiplets in that representation and $n_s$ be the number of singlets. We then have
\begin{equation}
\label{e-2-4}
n_H = \sum_k n_{i_1 \ldots i_k} \dim \MR_{i_1} \ldots \dim \MR_{i_k} + n_s.
\end{equation}
The full spectrum of the theory is thus given by
\begin{equation}
\label{e-2-5} ( g_{\mu\nu} , B^+_{\mu\nu} , \psi^{i-}_\mu) + n_T ( B^-_{\mu\nu} , \phi , \chi^{i+} ) + n_V ( A_\mu, \lambda^{i-} ) + n_H ( 4 \varphi , 2 \psi^{+} ).
\end{equation}
In general, the above spectrum is anomalous. The anomalies of the theory fall into two types. The first type corresponds to the usual local gravitational, gauge and mixed anomalies present in chiral theories. The second type pertains to the gauge sector of the theory and corresponds to the global anomalies arising from gauge transformations not continuously connected to the identity. Below, we shall examine the two cases in turn and we will state the necessary and sufficient conditions for the absence of the two types of anomalies.

\subsection{Local anomalies}
\label{sec-2-2}

Let us first examine the local anomalies. Starting from gravitational anomalies, we use the normalization of Appendix \ref{appa} to represent the total gravitational anomaly of the theory by the anomaly 8--form
\begin{equation}
\label{e-2-6}
I_{8} (R) = \frac{n_H - n_V + 29 n_T - 273}{360} \tr R^4 + \frac{n_H - n_V - 7 n_T + 51}{288} (\tr R^2)^2 .
\end{equation}
Passing to the gauge and mixed anomalies, we have to introduce some notation. We let $F_\alpha$ and $F_r$ be the field strengths associated with $\MG_\alpha$ and $\MG_r$. We also let $n_{\alpha,i}$ and $n_{\alpha\beta,ij}$ denote the numbers of hypers transforming in the representation $\MR_i$ of $\MG_\alpha$
and in $(\MR_i,\MR_j)$ of $\MG_\alpha \times \MG_\beta$. Then, using the formulas in Appendix A, we write the gauge anomaly polynomial as
\begin{eqnarray}
\label{e-2-7} I_8 (F) &=& - \frac{2}{3} \sum_\alpha \bigl( \Tr F_\alpha^4 - \sum_i n_{\alpha,i} \tr_i F_\alpha^4 \bigr) + 4 \sum_{\alpha < \beta} \sum_{i,j} n_{\alpha\beta,ij} \tr_i F_\alpha^2 \tr_j F_\beta^2 \nn\\ && - \frac{2}{3} \bigl[ \tr' F_r^4 + ( \dim \MG_s + \dim \MG_a + 5 - n_T ) \tr F_r^4 \bigr] \nn\\ &&- 4 \Tr F_\alpha^2 \tr' F_r^2 ,
\end{eqnarray}
where ``$\Tr$'' and ``$\tr_i$'' stand for the traces in the adjoint and $\MR_i$ of $\MG_\alpha$ while ``$\tr$'' and ``$\tr'$'' stand for the traces in the fundamental of $\MG_r$ and in the representation of the gauginos. The four terms in (\ref{e-2-7}) are recognized as (i) the contribution of the gauginos and hyperinos to the anomaly under the pure $\MG_\alpha$ factors, (ii) the contribution of the hyperinos to the anomaly under the
products  $\MG_\alpha \times \MG_\beta$, (iii) the contribution of the $\MG_r$, $\MG_s$ and $\MG_a$ gauginos, gravitino and tensorinos to the anomaly under $\MG_r$ and (iv) the contribution of the gauginos to the anomaly under $\MG_\alpha \times \MG_r$. In a similar manner, we find the mixed anomaly
\begin{eqnarray}
\label{e-2-8} I_8 (F,R) &=& \frac{1}{6} \tr R^2 \sum_\alpha \bigl( \Tr F_\alpha^2 - \sum_i n_{\alpha,i} \tr_i F_\alpha^2 \bigr) \nn\\ &&+ \frac{1}{6} \tr R^2 \bigl[ \tr' F_r^2 + ( \dim \MG_s + \dim \MG_a - 19 - n_T ) \tr F_r^2 \bigr].
\end{eqnarray}
Eqs. (\ref{e-2-7}) and (\ref{e-2-8}) can be brought into a more convenient form by expressing all traces in terms of a single representation, which we take to be the fundamental. For each $\MG_\alpha$ we will have expressions of the form
\begin{equation}
\label{e-2-9}
\tr_i F_\alpha^4 = a_{\alpha,i} \tr F_\alpha^4 + b_{\alpha,i} ( \tr F_\alpha^2 )^2 ,\qquad \tr_i
F_\alpha^2 = c_{\alpha,i} \tr F_\alpha^2,
\end{equation}
where the various group- and representation-dependent coefficients $a_{\alpha,i}$, $b_{\alpha,i}$ and $c_{\alpha,i}$ are given in Appendix \ref{appb}. Similarly, for $\MG_r$ we will have
\begin{eqnarray}
\label{e-2-10} &\tr' F_r^4 = b'_r ( \tr F_r^2 )^2 ,\qquad \tr F_r^4 = b_r ( \tr F_r^2 )^2, \nn\\ &  \tr' F_r^2 = c'_r \tr F_r^2,
\end{eqnarray}
where
\begin{eqnarray}
\label{e-2-11}
b'_r = \frac{83}{2} ,\qquad b_r = \frac{1}{2} ,\qquad c'_r = 11;&& \qquad \text{if }\MG_r = \textrm{Sp}(1),\nn\\
b'_r = b_{R} = c'_r = 1;&& \qquad \text{if }\MG_r =
\textrm{U}(1).
\end{eqnarray}
Using (\ref{e-2-9}) and (\ref{e-2-10}), introducing the quantities
\begin{eqnarray}
\label{e-2-12}
A_\alpha \equiv a_{\alpha,\MA} - \sum_i n_{\alpha,i} a_{\alpha,i}, &&
\nn\\
B_\alpha \equiv b_{\alpha,\MA} - \sum_i n_{\alpha,i} b_{\alpha,i}, &&\qquad B_r \equiv b'_r + ( \dim \MG_s + \dim \MG_a + 5 - n_T ) b_{R} , \nn\\
C_\alpha \equiv c_{\alpha,\MA} - \sum_i n_{\alpha,i} c_{\alpha,i}, &&\qquad C_r \equiv c'_r + \dim \MG_s + \dim \MG_a - 19 - n_T, \nn\\
C_{\alpha\beta} \equiv \sum_{i,j} n_{\alpha\beta,ij} c_{\alpha,i}
c_{\beta,j}, &&\qquad C_{\alpha,r} \equiv - c_{\alpha,\MA},
\end{eqnarray}
and employing the extended index $A$, we write the gauge and mixed anomaly polynomials of the theory in the compact forms
\begin{equation}
\label{e-2-13}
I_8(F) = - \frac{2}{3} \sum_\alpha A_\alpha \tr F_\alpha^4 - \frac{2}{3} \sum_A B_A (\tr F_A^2)^2 + 4 \sum_{A < B} C_{AB} \tr F_A^2 \tr F_B^2,
\end{equation}
and
\begin{equation}
\label{e-2-14}
I_8(F,R) = \frac{1}{6} \tr R^2 \sum_A C_A \tr F_A^2.
\end{equation}
Combining (\ref{e-2-6}), (\ref{e-2-13}) and (\ref{e-2-14}), we finally find the total anomaly polynomial
\begin{eqnarray}
\label{e-2-15}
I_8 &=& \frac{n_H - n_V + 29 n_T - 273}{360} \tr R^4 + \frac{n_H - n_V - 7 n_T + 51}{288} (\tr R^2)^2 \nn\\ &&+ \frac{1}{6} \tr R^2 \sum_A C_A \tr F_A^2 \nn\\ &&- \frac{2}{3} \sum_\alpha A_\alpha \tr F_\alpha^4 - \frac{2}{3} \sum_A B_A (\tr F_A^2)^2 + 4 \sum_{A < B} C_{AB} \tr F_A^2 \tr F_B^2.
\end{eqnarray}
If the total anomaly is to cancel through a Green-Schwarz--type mechanism, the above polynomial must factorize. A necessary condition for this is that all irreducible $\tr R^4$ and $\tr F_\alpha^4$ terms in (\ref{e-2-15}) must vanish. Regarding the $\tr R^4$ term, the fact that $\textrm{SO}(5,1)$ possesses a fourth-order Casimir implies that the coefficient of this term must vanish. This way, we are led to our first constraint
\begin{equation}
\label{e-2-16}
n_H - n_V = 273 - 29 n_T,
\end{equation}
which clearly shows that the presence of hypermultiplets is necessary for anomaly cancellation at least for $n_T \leqslant 9$. Passing to the $\tr F_\alpha^4$ terms, their vanishing requires that
\begin{equation}
\label{e-2-17}
A_\alpha = 0 ;\qquad \textrm{for all } \alpha.
\end{equation}
This can be achieved either (i) if the relevant representations of $\MG_\alpha$ have no fourth-order invariants ($a_{\alpha,i}=0$ for all $i$) or (ii) if the $n_{\alpha,i}$'s are chosen appropriately. Provided that (\ref{e-2-16}) and (\ref{e-2-17}) hold, the anomaly polynomial reads
\begin{equation}
\label{e-2-18} I_8 = K (\tr R^2)^2 + \frac{1}{6} \tr R^2 \sum_A C_A \tr F_A^2 - \frac{2}{3} \sum_A B_A (\tr F_A^2)^2 + 4 \sum_{A < B} C_{AB} \tr F_A^2 \tr F_B^2.
\end{equation}
where we introduced the quantity
\begin{equation}
\label{e-2-19} K = \frac{9 - n_T}{8}.
\end{equation}

To make a general analysis of the factorization properties of this polynomial, it helps to treat the Lorentz group in an equal footing with the other gauge groups by defining $F_0 = R$ as in \cite{Randjbar-Daemi:1985wc}. Introducing a summation index $I=0,1,\ldots,N$, we can then represent the anomaly polynomial in the concise form
\begin{equation}
\label{e-2-20} I_8 = G^{IJ} \tr F_I^2 \tr F_J^2,
\end{equation}
where $\MBFG$ is a real symmetric $(N+1) \times (N+1)$ matrix with entries
\begin{equation}
\label{e-2-21}
G^{00} = K ,\qquad G^{0A} = \frac{C_A}{12} ,\qquad G^{AA} = - \frac{2 B_A}{3} ,\qquad G^{AB} = 2 C_{AB} \text{ ($A \ne B$)}.
\end{equation}
The anomaly cancellation conditions depend on the properties of the matrix $\MBFG$ as well as on the number $n_T$ of available tensor multiplets. The two possible mechanisms are as follows.
\begin{enumerate}
\item {\bf Green-Schwarz mechanism.} For an arbitrary number of tensor multiplets\footnote{In discussing the $n_T \ne 1$ case, we ignore subtleties related to the construction of actions for (anti-)self-dual 2--forms.}, anomalies may be cancelled by the standard Green-Schwarz mechanism. In order for that mechanism to be applicable, the matrix $\MBFG$ must be a matrix of rank $r \leqslant 2$; if $r=2$, the (real) nonzero eigenvalues $\lambda_m$, $m=0,1$, must satisfy $\lambda_0 \lambda_1 < 0$. For $r=2$, we may define $c^{mI}$ as the eigenvectors corresponding to $\lambda_m$ multiplied by $|\lambda_m|^{1/2}$ and write the similarity transformation of $\MBFG$ in the form
\begin{equation}
\label{e-2-22}
G^{IJ} = \epsilon \eta_{mn} c^{mI} c^{nJ} = \frac{1}{2} ( u^I v^J + v^I u^J ),
\end{equation}
where $\epsilon$ is the sign of $\lambda_0$, $\eta_{mn}=\diag(1,-1)$ is the $\textrm{SO}(1,1)$--invariant tensor and
\begin{equation}
\label{e-2-23}
u^I \equiv \epsilon( c^{0I} - c^{1I} ) ,\qquad v^I \equiv c^{0I} + c^{1I}.
\end{equation}
Using (\ref{e-2-22}), we can write the anomaly polynomial in the factorized form
\begin{equation}
\label{e-2-24}
I_8 = \epsilon \eta_{mn} c^{mI} c^{nJ} \tr F_I^2 \tr F_J^2 = u^I \tr F_I^2 v^J \tr F_J^2.
\end{equation}
This anomaly can cancel by the standard Green-Schwarz mechanism. Letting $B_2^{(0)} = B_2^+$ be the self-dual 2--form of the gravity multiplet and $B_2^{(1)}$ be any one of the anti-self-dual 2--forms in the tensor multiplets and setting $B_2 = B_2^{(0)} + B_2^{(1)}$, we construct the Green-Schwarz term
\begin{equation}
\label{e-2-25}
S_{GS} \sim \int u^I B_2 \tr F_I^2,
\end{equation}
and we modify the gauge/Lorentz transformation law of the $B_2$'s to
\begin{equation}
\label{e-2-26}
\delta B_2 \sim v^I \omega^1_{2,I},
\end{equation}
where $\omega^1_{2,I}$ is related to $\tr F_I^2$ by descent. The variation of (\ref{e-2-25}) under (\ref{e-2-26}) exactly cancels the anomaly of the theory. For $r=1$, one may repeat the above discussion with the
appropriate $c^{mI}$ set to zero.

\item {\bf Generalized Green-Schwarz mechanism.} In the case $n_T > 1$, there exists a generalization of the Green-Schwarz mechanism, found by Sagnotti \cite{Sagnotti:1992qw}, which allows for anomaly cancellation under weaker constraints. For that mechanism to apply, the matrix $\MBFG$ must be a matrix of rank $r \leqslant
n_T+1$ whose nonzero eigenvalues $\lambda_m$, $m=0,\ldots,r-1$, include an eigenvalue $\lambda_0$ such that $\lambda_0 \lambda_m < 0$ for $m > 0$. For $r=n_T+1$, we may define $c^{mI}$ as before and we write the similarity transformation of $\MBFG$ as
\begin{equation}
\label{e-2-27}
G^{IJ} = \epsilon \eta_{mn} c^{mI} c^{nJ} = \frac{1}{2} \sum_{i=1}^{n_T} ( u^{iI} v^{iJ} + v^{iI} u^{iJ} ),
\end{equation}
where now $\eta_{mn}=\diag(1,-1,\ldots,-1)$ is the $SO(1,n_T)$--invariant metric and
\begin{eqnarray}
\label{e-2-28}
u^{iI} \equiv \epsilon\left( \frac{c^{0I}}{\sqrt{n_T}} - c^{iI} \right) ,\qquad v^{iI} \equiv \frac{c^{0I}}{\sqrt{n_T}} + c^{iI}.
\end{eqnarray}
This way, the anomaly polynomial is written as a sum of factorized terms,
\begin{equation}
\label{e-2-29}
I_8 = \epsilon \eta_{mn} c^{mI} c^{nJ} \tr F_I^2 \tr F_J^2 = \sum_{i=1}^{n_T} u^{iI} v^{iJ} \tr F_I^2 \tr F_J^2.
\end{equation}
This anomaly can cancel by a generalization of the Green-Schwarz mechanism. Letting $B_2^{(0)} = B_2^+$ and $B_2^{(i)}$ be the anti-self-dual 2--forms in the tensor multiplets, we construct the $\textrm{SO}(1,n_T)$--invariant generalized Green-Schwarz term \cite{Sagnotti:1992qw,Riccioni:1998th}
\begin{equation}
\label{e-2-30}
S_{GS} \sim \int \epsilon \eta_{mn} c^{mI} B_2^{(n)} \tr F_I^2,
\end{equation}
and we modify the gauge/Lorentz transformation law of the $B_2$'s to
\begin{equation}
\label{e-2-31}
\delta B_2^{(m)} \sim c^{mI} \omega^1_{2,I}.
\end{equation}
Again, for $r<n_T+1$, one may repeat the above discussion with the appropriate $c^{mI}$'s set to zero.
\end{enumerate}

In this paper, we will only consider theories whose anomalies cancel by the standard Green-Schwarz mechanism. To examine the conditions for anomaly cancellation, it is very useful to state them in a more explicit form. To do so, we compare the general form (\ref{e-2-21}) of $\MBFG$ with the expression (\ref{e-2-22}). Comparison of the $G^{00}$, $G^{0A}$, $G^{AA}$ and $G^{AB}$ terms leads respectively to the conditions
\begin{eqnarray}
\label{e-2-32} &u^0 v^0 = K, \\
\label{e-2-33} &u^0 v^A + v^0 u^A = \frac{C_A}{6} ,\qquad u^A v^A
= - \frac{2 B_A}{3}; \qquad &\textrm{for all } A,\\
\label{e-2-34} &u^A v^B + v^A u^B = 4 C_{AB} ;\qquad &\textrm{for
all } A < B.
\end{eqnarray}
To begin, we note that we can set $u^0 = K$ and $v^0 = 1$ without loss of generality. To proceed, we have to distinguish between the cases $n_T\ne9$ ($K \ne 0$) and $n_T=9$ ($K=0$).
\begin{itemize}
\item $n_T \ne 9$. In this case, Eqs. (\ref{e-2-33}) imply that $u^A$ and $Kv^A$ must be roots of the equation
\begin{equation}
\label{e-2-35}
x^2 - \frac{C_A}{6} x - \frac{2 K B_A}{3} = 0,
\end{equation}
and, in order for them to be real, we must have
\begin{equation}
\label{e-2-36}
C_A^2 + 96 K B_A \geqslant 0;\qquad \textrm{for all } A.
\end{equation}
Finally, Eq. (\ref{e-2-34}) leads to the condition
\begin{equation}
\label{e-2-37}
C_A C_B \pm \sqrt{ ( C_A^2 + 96 K B_A )( C_B^2 + 96 K B_B )} = 288 K C_{AB},
\end{equation}
for at least one choice for the $u_A$'s and $Kv_A$'s as roots of (\ref{e-2-35}), the plus or minus sign depending on the particular choice. E.g. in the case of three groups, (\ref{e-2-34}) is satisfied when (\ref{e-2-37}) holds for each pair $AB =(12,13,23)$ with either one of the sign combinations $(-,-,-)$, $(-,+,+)$,
$(+,-,+)$ and $(+,+,-)$.

\item $n_T = 9$. In that case, the first of Eqs. (\ref{e-2-33}) determines $u^A = C_A/6$, the second of Eqs. (\ref{e-2-33}) gives
\begin{equation}
\label{e-2-38}
C_A v^A = - 4 B_A;\qquad \textrm{for all } A,
\end{equation}
and Eq. (\ref{e-2-34}) gives
\begin{equation}
\label{e-2-39}
C_A v^B + C_B v^A = 24 C_{AB};\qquad \textrm{for all } A < B.
\end{equation}
Eqs. (\ref{e-2-38}) and (\ref{e-2-39}) together form an overdetermined linear system of $N(N+1)/2$ equations for $N$ unknowns. In the general case, the system has the form $\MBFA \mbfv = \mbfb$ with
\begin{equation}
\label{e-2-40} {\scriptsize \MBFA = \left(
\begin{array}{ccccc}
C_1 & 0 & \cdots & 0 & 0 \\
0 & C_2 & \cdots & 0 & 0 \\
\vdots & \vdots & \ddots & \vdots & \vdots \\
0 & 0 & \cdots & C_{N-1} & 0 \\
0 & 0 & \cdots & 0 & C_N \\
C_2 & C_1 & \ldots & 0 & 0 \\
\vdots & \vdots & \ddots & \vdots & \vdots \\
C_{N-1} & 0 & \ldots & C_1 & 0 \\
C_N & 0 & \ldots & 0 & C_1 \\
\vdots & \ddots &  & \vdots & \vdots \\
\vdots & \vdots & \ddots & \vdots & \vdots \\
0 & 0 & \ldots & C_N & C_{N-1}
\end{array} \right),\quad
\mbfv = \left(\begin{array}{c}
v^1 \\
v^2 \\
\vdots\\
v^{N-1} \\
v^N
\end{array} \right) ,\quad
\mbfb = \left(
\begin{array}{c}
-4 B_1 \\
-4 B_2 \\
\vdots\\
-4 B_{N-1} \\
-4 B_N \\
24 C_{12} \\
\vdots\\
24 C_{1,N-1} \\
24 C_{1N} \\
\vdots\\
\vdots\\
24 C_{N-1,N}
\end{array} \right)},
\end{equation}
and the constraints determining whether it has solutions are given by
\begin{equation}
\label{e-2-41} \det \MBFC = 0; \qquad \text{ for every $(N+1)\times(N+1)$ submatrix $\MBFC$ of $(\MBFA,\mbfb)$}.
\end{equation}
From (\ref{e-2-40}), we can see that when we have $B_A=C_A=0$ for all $A \ne \bar{A}$ and $C_{AB}=C_{BA}=0$ for $A,B \ne \bar{A}$ where $\bar{A}$ is a given value of the index $A$, the system reduces to $N$ independent equations and has always a solution. For Poincar\'e supergravities, this corresponds to the case where the hypermultiplets transform in the adjoint representation of $N-1$ group factors in $\MG_s$ and in an arbitrary representation of the remaining factor. For gauged supergravities, this corresponds to the case where the hypermultiplets transform in the adjoint of $\MG_s$; these were the solutions found in \cite{Salam:1985mi}.
\end{itemize}
To summarize, the requirement of Green-Schwarz cancellation of local anomalies in $D=6$, $\MN=1$ supergravity boils down (for $n_T \ne 9$) to the four conditions (\ref{e-2-16}), (\ref{e-2-17}), (\ref{e-2-36}) and (\ref{e-2-37}). The first condition fixes the number of hypermultiplets in terms of the gauge group. The second condition either holds identically (in the absence of fourth-order Casimirs) or constrains the numbers of representations (in the presence of fourth-order Casimirs). The third condition is an inequality whose main effect is to forbid higher representations (for which $B_\alpha$ can attain large negative values). Finally, the fourth condition imposes a very stringent constraint on the numbers of representations; it is this latter condition that seriously reduces the number of possible models in the case when product groups are considered. In the special case $n_T=9$, the last two conditions are replaced by (\ref{e-2-41}).

\subsection{Global anomalies}
\label{sec-2-3}

Besides the perturbative anomalies described above, there is also the possibility that the theory may suffer from global anomalies of the type first discovered by Witten \cite{Witten:1982fp} in the context of a 4D $\textrm{SU}(2)$ gauge theory. In our 6D case \cite{Elitzur:1984kr,Kiritsis:1986mf}, such anomalies may arise if the sixth homotopy group $\pi_6 (\MG)$ of the gauge group is non-trivial. If that is the case, the space of gauge transformations is disconnected and so there exist ``large'' gauge transformations not connected to the identity. Under such transformations, the fermion determinant may pick up a phase factor and is therefore ill-defined unless the numbers of fermions are such that this factor equals unity. This requirement provides additional constraints on the spectrum of the theory.

The only simple groups with non-trivial sixth homotopy groups are $G_2$, $\textrm{SU}(3)$ and $\textrm{SU}(2)$. For these groups,
\begin{equation}
\label{e-2-42}
\pi_6 (G_2) = \MBBZ_3 ,\quad\quad \pi_6 \left( \textrm{SU}(3) \right) = \MBBZ_6 ,\quad\quad \pi_6 \left(
\textrm{SU}(2) \right) = \MBBZ_{12}.
\end{equation}
The conditions for the absence of global anomalies in the presence of a factor $\MG_\alpha=G_2,\textrm{SU}(3),\textrm{SU}(2)$ in the $\MG_s$ part of the gauge group can be found in \cite{Bershadsky:1997sb} and they amount to the following integrality constraints
\begin{align}
\label{e-2-43}
&\MG_\alpha=G_2 :            &&1 - 4 \sum_i n_{\alpha,i} b_{\alpha,i} = 0 \mod 3, \nn\\
&\MG_\alpha=\textrm{SU}(3) : &&- 2 \sum_i n_{\alpha,i} b_{\alpha,i} = 0 \mod 6, \nn\\
&\MG_\alpha=\textrm{SU}(2) : &&4 - 2 \sum_i n_{\alpha,i} b_{\alpha,i} = 0 \mod 6.
\end{align}
where $n_{\alpha,i}$ and $b_{\alpha,i}$ are defined in \S\ref{sec-2-1}. Note that, when the whole $\textrm{Sp}(1)_R \cong \textrm{SU}(2)_R$ is gauged, there are also global R-symmetry anomalies. The condition for their absence is given by
\begin{align}
\label{e-2-44}
&\MG_r=\textrm{Sp}(1) : && 4 + \dim \MG_s + \dim \MG_a - n_T = 0 \mod 6.
\end{align}
Eqs. (\ref{e-2-43}) and (\ref{e-2-44}) must be solved together with the local anomaly cancellation conditions of the previous subsection in order to determine the possible anomaly-free models.

\subsection{Searching for anomaly-free theories}
\label{sec-2-4}

The purpose of this paper is to conduct a systematic search for 6D supergravity models satisfying the anomaly cancellation conditions stated above. Since a complete classification seems to be very difficult, we will make several assumptions, expected to hold for many models of potential physical interest. The restrictions to be imposed are the following.
\begin{enumerate}
\item The theory contains only one tensor multiplet, $n_T=1$.
\item The semisimple gauge group factor $\MG_s$ is a product of up to two simple groups.
\item The hypermultiplets may transform in a set of low-dimensional representations of the simple factors in $\MG_s$. The representations to be considered are shown on Table \ref{t-1}.
\item For Poincar\'e theories, the allowed exceptional groups are $E_8$, $E_7$, $E_6$ and $F_4$ and the allowed classical groups are $\textrm{SU}(5 \leqslant N \leqslant 32)$, $\textrm{SO}(10 \leqslant N \leqslant 64)$ and $\textrm{Sp}(4 \leqslant N \leqslant 32)$. At most one simple factor in $\MG_s$ may be a classical group. The abelian factor $\MG_a$ is empty.
\item For gauged theories, all exceptional groups are allowed while the allowed classical groups are as before. At most one simple factor in $\MG_s$ may be a classical group. The abelian factor $\MG_a$ can be non-trivial.
\end{enumerate}
\begin{table}[!t]
\begin{center}
\begin{tabular}{|c||l|l|}
\hline
Group & Low-dimensional Irreps & Comments\\
\hline
\hline
$E_8$ & $\mathbf{248}$ &\\
\hline
$E_7$ & $\mathbf{56}^*,\mathbf{133},\mathbf{912}^*$ & ${}^*$pseudoreal \\
\hline
$E_6$ & $\mathbf{27},\mathbf{78},\mathbf{351},\mathbf{351'},\mathbf{650}$ &\\
\hline
$F_4$ & $\mathbf{26},\mathbf{52},\mathbf{273},\mathbf{324}$ &\\
\hline
$G_2$ & $\mathbf{7},\mathbf{14},\mathbf{27},\mathbf{64}$ &\\
\hline
\hline
$\textrm{SU}(N)$ & $\mathbf{N},\mathbf{N^2-1},\mathbf{\frac{N(N-1)}{2}},\textstyle{\mathbf{\frac{N(N+1)}{2}}}$ &\\
\hline
$\textrm{SO}(N)$ & $\mathbf{N}, \mathbf{\frac{N(N-1)}{2}}, \mathbf{2^{\lfloor \frac{N+1}{2} \rfloor - 1}}^*$ & ${}^*$pseudoreal if $N = 3,4,5 \mod 8$ \\
\hline
$\textrm{Sp}(N)$ & $\mathbf{2N}^*,\mathbf{N(2N+1)},\mathbf{N(2N-1)-1}$ & ${}^*$pseudoreal \\
\hline
\end{tabular}
\end{center}
\caption{The possible simple gauge groups and their low-dimensional representations.}
\label{t-1}
\end{table}
All of these assumptions have been made on a purely practical basis. In particular, the lower bounds on the group rank as well as the restriction to at most one classical group factor were imposed because the proliferation of possible models in the case these assumptions were relaxed would make the exhaustive search for anomaly-free models and their classification an intractable task.

In the next two sections, we will present the complete lists of anomaly-free models under these conditions, starting from the case of Poincar\'e supergravities and proceeding to the case of gauged supergravities. In the course, we will identify as many of the known models as possible and we will comment on their construction, their origin and their properties. The results presented should be read according to the convention that each representation, designated by its dimension, corresponds to all representations with the same dimension and second and fourth indices, i.e. to all representations related by symmetries such as complex conjugation and triality. Accordingly, the corresponding numbers of multiplets for a representation are understood as the total numbers of multiplets
in these representations. For example, in the case of $E_6$, the notation $\mathbf{27}$ refers to the two conjugate representations $\mathbf{27}$ and $\overline{\mathbf{27}}$ and the field content $n \cdot \mathbf{27}$ is understood as all combinations of the form $n_1 \cdot \mathbf{27} + n_2 \cdot \overline{\mathbf{27}}$ with $n_1 + n_2 = n$. Also, the numbers of singlet hypermultiplets for each model will not be displayed explicitly.

Finally, there are two issues referring to the reality properties of the representations under consideration. First, when there appear pseudoreal representations, one may allow the corresponding numbers of hypermultiplets to take half-integer values as well. For example, in the case of $E_7$, the notation $\frac{1}{2} \cdot \mathbf{56}$ refers to ``half'' a hypermultiplet in the pseudoreal representation $\mathbf{56}$, also understood as one hypermultiplet in the minimal representation $\mathbf{28}$. Second, in the case where there appear complex representations, CPT invariance requires that these representations occur in complex-conjugate pairs; it is only these representations that will be considered here.

\section{Anomaly-free Poincar\'e supergravities}
\label{sec-3}

In this section, we begin our search by considering the case of Poincar\'e supergravities, i.e. the case where the gauge group does not involve an R-symmetry subgroup. As mentioned in the introduction, the number of these models is expected to be quite large; it turns out that this is indeed the case. In the course of the search, we recover various known models already found in the literature, and we find some models not previously identified.

\subsection{Simple groups}
\label{sec-3-1}

Let us start from the case of one simple gauge group. In this case, the conditions to be solved are Eq. (\ref{e-2-16}) for the cancellation of the irreducible gravitational anomaly, Eq. (\ref{e-2-17}) for the cancellation of the irreducible gauge anomaly (when applicable) and the factorization condition (\ref{e-2-36}). Below, we present all possible models satisfying these conditions under the assumptions introduced at the end of Section 2. To make the discussion more pedagogical, we illustrate the procedure in detail.

For the exceptional groups, the only conditions to be solved are Eqs. (\ref{e-2-16}) and (\ref{e-2-36}). Using (\ref{e-2-3}) and (\ref{e-2-4}) and noting that the number of singlets must be a nonnegative integer, we see that the first condition constrains the number of charged hypermultiplets according to
\begin{equation}
\label{e-3-1}
\sum_i n_{i} \dim \MR_i \leqslant \dim \MG + 244.
\end{equation}
Also, using (\ref{e-2-13}), the second condition takes the explicit form
\begin{equation}
\label{e-3-2}
(c_{\MA} - \sum_i n_i c_i )^2 + 96 (b_{\MA} - \sum_i n_i b_i) \geqslant 0,
\end{equation}
where the subscript ``$\MA$'' refers to the adjoint. Also, since $G_2$, $\textrm{SU}(3)$ and $\textrm{SU}(2)$ are excluded from the search, we need not examine global anomalies. One can immediately see that Eqs. (\ref{e-3-1}) and (\ref{e-3-2}) are automatically satisfied when there is a hypermultiplet in the adjoint plus $244$ singlets or when all hypermultiplets are singlets; such solutions will be considered as trivial and will not be displayed. Our results are shown below.

\begin{enumerate}
\item $E_8$. For the $E_8$ gauge group we must have $n_H = 248 + 244 = 492$ and the only available representation is the adjoint. Since the hypermultiplets can fit in at most one adjoint, the only solutions are the trivial ones.
\item $E_7$. Since this is the first non-trivial case to be considered, we will present it in some detail. For the $E_7$ gauge group we must have $n_H = 133 + 244 = 377$ and the available representations are the adjoint $\mathbf{133}$ and the pseudoreal fundamental $\mathbf{56}$. So, the condition (\ref{e-3-1}) translates to
\begin{equation}
133 n_{\mathbf{133}} + 56 n_{\mathbf{56}} \leqslant 377,
\end{equation}
and is satisfied by the following matter content
\begin{align*}
&\textstyle{\frac{n}{2}} \cdot \mathbf{56} ; && n=0,\ldots,13, \\
&\mathbf{133} + \textstyle{\frac{n}{2}} \cdot \mathbf{56} ; && n=0,\ldots,8 \\
&2 \cdot \mathbf{133} + \textstyle{\frac{n}{2}} \cdot \mathbf{56}; &&
n=0,\ldots,3,
\end{align*}
plus the appropriate numbers of singlets. However, the second condition (\ref{e-3-2}), namely
\begin{equation} ( 3 - 3 n_{\mathbf{133}} - n_{\mathbf{56}} )^2 + 4 ( 4 - 4 n_{\mathbf{133}} - n_{\mathbf{56}}
) \geqslant 0,
\end{equation}
further restricts the possible solutions to
\begin{align}
&\textrm{(a) }\textstyle{\frac{n}{2}} \cdot \mathbf{56} ; && n=0,\ldots,13, \nn\\
&\textrm{(b) }\mathbf{133} + 4 \cdot \mathbf{56}.
\end{align}
Regarding the models (a), one may make a shift of $n$ to $n_1 = n+4$ and rewrite them as $\textstyle{\frac{n_1-4}{2}} \cdot \mathbf{56}$. These models are then recognized as those resulting from the $E_8 \times E_8$ heterotic string on $K3$ by embeddding $n_1$ units of instanton charge in an $\textrm{SU}(2)$ subgroup of the first $E_8$ (and ignoring the other $E_8$). These theories are the first ones in a chain of theories related to each other by successive Higgsing; in terms of theories to be discussed here, the relevant parts of the chain are $E_7\textrm{(a)} \to E_6\textrm{(a)} \to F_4\textrm{(a)}\to \ldots$ and $E_7\textrm{(a)} \to \textrm{SO}(11) \textrm{(b)} \to \textrm{SO}(10) \textrm{(b)} \to \ldots$.
\item $E_6$. Now, we have $n_H=322$ and the available representations are $\mathbf{27}$ and $\mathbf{78}$. Proceeding as before, we find the solutions
\begin{align}
&\textrm{(a) }2 n \cdot \mathbf{27} ;&& n=1,\ldots,5, \nn\\
&\textrm{(b) }4 \cdot \mathbf{78}, \nn\\
&\textrm{(c) }\mathbf{78} + 8 \cdot \mathbf{27}, \nn\\
&\textrm{(d) }2 \cdot \mathbf{78} + 6 \cdot \mathbf{27},
\end{align}
where, in addition, we imposed the requirement of CPT invariance which demands an even number of $\mathbf{27}$'s, understood as $2 n \cdot \mathbf{27} \to n \cdot \mathbf{27} + n \cdot \overline{\mathbf{27}}$.
\item $F_4$. Now, we have $n_H=296$ and the available representations are $\mathbf{26}$, $\mathbf{52}$ and
$\mathbf{273}$. The possible solutions are
\begin{align}
&\textrm{(a) }n \cdot \mathbf{26}; && n=0,\ldots,11, \nn\\
&\textrm{(b) }\mathbf{52} + 8 \cdot \mathbf{26}, \nn\\
&\textrm{(c) }n \cdot \mathbf{52} + (11-2n) \cdot \mathbf{26}; &&
n=1,\ldots,5.
\end{align}
\end{enumerate}

For the classical groups, there is the extra condition (\ref{e-2-17}) which we write explicitly as
\begin{equation}
\sum_i n_i a_i = a_{\MA}
\end{equation}
Again, there exist trivial solutions, corresponding to a hypermultiplet in the adjoint plus $244$ singlets, that will not be displayed. The search for anomaly-free models can be conducted as before and the results are summarized as follows.

\begin{enumerate}

\item $\textrm{SU}(N)$:
\begin{align}
&5 \leqslant N \leqslant 18: &&\textrm{(a) }\left[ 2N - 2n (N-8) \right] \cdot \mathbf{N} + 2n \cdot \textstyle{\mathbf{\frac{N(N-1)}{2}}}. \\
&N=8: &&\textrm{(b) }\mathbf{63} + 8 \cdot \mathbf{28}.\\
&N=7: &&\textrm{(b) }\mathbf{48} + 8 \cdot \mathbf{7} + 8 \cdot \mathbf{21}.\\
&N=6: && \textrm{(b) } \mathbf{35} + 16 \cdot \mathbf{6} + 8 \cdot \mathbf{15}, \nn\\
      &&&\textrm{(c) } 2 \cdot \mathbf{35} + 8 \cdot \mathbf{6} + 10 \cdot \mathbf{15},\nn\\
      &&&\textrm{(d) }8 \cdot \mathbf{6} + 2 \cdot \mathbf{21} + 12 \cdot \mathbf{15}.\\
&N=5: &&  \textrm{(b) }\mathbf{24} + 24 \cdot \mathbf{5} + 8 \cdot \mathbf{10},\nn\\
      &&& \textrm{(c) }2 \cdot \mathbf{24} + 20 \cdot \mathbf{5} + 10 \cdot \mathbf{10},\nn\\
      &&& \textrm{(d) }4 \cdot \mathbf{24} + 6 \cdot \mathbf{5} + 12 \cdot \mathbf{10},\nn\\
      &&& \textrm{(e) }2 \cdot \mathbf{24} + 6 \cdot \mathbf{5} + 2 \cdot \mathbf{15} + 14 \cdot \mathbf{10},\nn\\
      &&& \textrm{(f) }6 \cdot \mathbf{5} + 4 \cdot \mathbf{15} + 14 \cdot \mathbf{10},\nn\\
      &&& \textrm{(g) }20 \cdot \mathbf{5} + 2 \cdot \mathbf{15} + 12 \cdot \mathbf{10}.
\end{align}
In the first series of solutions, $n$ is restricted to all integer values such that all multiplicities, including the $243 - N^2 + n \textstyle{\frac{N^2 - 15N}{2}}$ singlets, are nonnegative.
\item $\textrm{SO}(N)$:
\begin{align}
&\!\!\!\!\!\!\!\! 10 \leqslant N \leqslant 30: &&\textrm{(a) } (N-8) \cdot \mathbf{N}. \\
&\!\!\!\!\!\!\!\! 10 \leqslant N \leqslant 14: &&\textrm{(b) } \textstyle{\mathbf{\frac{N(N-1)}{2}}} + 8 \cdot \mathbf{N} +
2^{8 - \lfloor \frac{N+1}{2} \rfloor} \cdot \mathbf{2^{\lfloor \frac{N+1}{2} \rfloor - 1}}. \\
&\!\!\!\!\!\!\!\! N=14: && \textrm{(c) } (4n+6) \cdot \mathbf{14} + n \cdot \mathbf{64}; && n=1,2. \\
&\!\!\!\!\!\!\!\! N=13: && \textrm{(c) } (2n+5) \cdot \mathbf{13} + \textstyle{\frac{n}{2}} \cdot \mathbf{64}; && n=1,4. \\
&\!\!\!\!\!\!\!\! N=12: && \textrm{(c) } (n+4) \cdot \mathbf{12} + \textstyle{\frac{n}{2}} \cdot \mathbf{32}; && n=1,\ldots,9, \nn\\
       &&&\textrm{(d) } 2 \cdot \mathbf{66} + 4 \cdot \mathbf{12} + 4 \cdot \mathbf{32}. \\
&\!\!\!\!\!\!\!\! N=11: && \textrm{(c) } (n+3) \cdot \mathbf{11} + \textstyle{\frac{n}{2}} \cdot \mathbf{32}; && n=1,\ldots,9, \nn\\
       &&&\textrm{(d) } n \cdot \mathbf{55} + (13-4n) \cdot \mathbf{11} + \textstyle{\frac{10-n}{2}} \cdot                        \mathbf{32}; && n=1,\ldots,3. \\
&\!\!\!\!\!\!\!\! N=10: && \textrm{(c) } (n+2) \cdot \mathbf{10} + n \cdot \mathbf{16}; && n=1,\ldots,10, \nn\\
       &&&\textrm{(d) } n \cdot \mathbf{45} + (12-3n) \cdot \mathbf{10} + (10-n) \cdot \mathbf{16}; && n=1,\ldots,4.
\end{align}
Let us try to identify some known models.

\begin{itemize}
\item In the first series of models, the $\textrm{SO}(28)\textrm{(a)}$ model is identified with the theory obtained from the $\textrm{SO}(32)$ heterotic string on $K3$ by embedding all 24 units of $K3$ instanton charge into one of the $\textrm{SU}(2)$ factors in the decomposition $\textrm{SO}(32) \supset \textrm{SO}(28) \times \textrm{SU}(2) \times \textrm{SU}(2)$ and breaking the other $\textrm{SU}(2)$ factor by Higgsing. By further Higgsing of this theory, one obtains all the $N < 28$ theories. Note that our list also contains models for $N=29,30$ which cannot be realized in a compactification context.

\item The $\textrm{SO}(12)\textrm{(c)}$ models are identified with the theories resulting from the $E_8 \times E_8$ heterotic string on $K3$, this time by embedding $n+4$ units of instanton charge in an $\textrm{SU}(2) \times \textrm{SU}(2)$ subgroup of the first $E_8$. These theories are also the first ones in a Higgs chain; in terms of the theories to be discussed here, the relevant parts of the chain are $\textrm{SO}(12)\textrm{(c)} \to \textrm{SO}(11) \textrm{(c)} \to \textrm{SO}(10) \textrm{(c)} \to \ldots$ and $\textrm{SO}(12)\textrm{(c)} \to \textrm{SU}(6) \textrm{(a)} \to \textrm{SU}(5) \textrm{(a)} \to \ldots$. The $E_7\textrm{(a)}$ and $\textrm{SO}(12) \textrm{(c)}$ models together form the top of the ``Higgs tree'' that contains all possible chains of theories that can be obtained from them by Higgsing. All these chains were constructed in \cite{Bershadsky:1996nh} by geometric engineering via F-theory.

\item The $\textrm{SO}(13)\textrm{(c)}$ models can also be realized \cite{Bershadsky:1997sb} from the $E_8 \times E_8$ heterotic string on $K3$ by considering the decomposition $E_8 \supset \textrm{SO}(16) \supset \textrm{SO}(13) \times \textrm{SU}(2)$ and embedding $n+4$ units of instanton charge in $\textrm{SU}(2)$.
\end{itemize}

\item $\textrm{Sp}(N)$:
\begin{align}
&\!\!\!\!\!\!\!\!\!\!\!\!\!\!\!\! 4 \leqslant N \leqslant 9: && \textrm{(a) } \left[ (2N+8) - n(2N-8) \right] \cdot \mathbf{2 N} + n \cdot \mathbf{N(2N-1)-1},\\
&\!\!\!\!\!\!\!\!\!\!\!\!\!\!\!\! N=4: &&  \textrm{(b) }\mathbf{36} + (n+8) \cdot \mathbf{27}; && n=0,1.
\end{align}
In the first series of solutions, $n$ is restricted to all integer values such that all multiplicities, including the $244 - 4 N^2 - 16N + n(6 N^2 - 17 N - 1)$ singlets, are nonnegative.

The first series of models has been identified in the literature \cite{Bershadsky:1996nh} as models with perturbatively enhanced symmetry resulting from F-theory compactifications on elliptic Calabi-Yau 3-folds based on the Hirzebruch surface; in this description, they originate from an $A_{2N-1}$ singularity on the coordinate of the $\mathbb{CP}^1$ fiber in that surface. The cases $n=0,1$ in this series were also given a gauge-theory interpretation \cite{Intriligator:1997kq} in terms of Type I D5--branes ($\textrm{SO}(32)$ small instantons) placed at a $\MBBZ_2$ orbifold singularity. For $n=1$, where the field content is given by
\begin{align}
&16 \cdot \mathbf{2N} + \mathbf{N(2N-1)-1},
\end{align}
the theory is on the Higgs branch. For $n=0$, where the field content is
\begin{align}
&(2N+8) \cdot \mathbf{2N},
\end{align}
the positions of all instantons are fixed, the blowing-up mode is zero and the theory rests on a non-trivial RG fixed point at the origin of the Coulomb branch.

\end{enumerate}

\subsection{Products of two simple groups}
\label{sec-3-2}

We now pass to the more complicated task of identifying anomaly-free models where the gauge group contains two simple group factors, $\MG = \MG_1 \times \MG_2$. This time, Eq. (\ref{e-2-17}) (when applicable) and Eq. (\ref{e-2-36}) must hold for each one of $\MG_1$ and $\MG_2$, while we also have the strict equality (\ref{e-2-37}) involving both group factors. Before we begin, we note that each of the simple-group solutions for, say, $\MG_1$ can be extended to a solution for $\MG_1 \times \MG_2$ by simply adding one adjoint of $\MG_2$. Such ``reducible'' solutions will not be written out explicitly.

We start our search from the case where both $\MG_1$ and $\MG_2$ are exceptional groups, in which case there are no fourth-order Casimirs. The largest-rank group of this type is $E_8 \times E_8$, which is one of the possible gauge groups of heterotic string theory; it is easily seen that this group admits only the trivial solutions. The group $E_8 \times E_7$ ($E_7 \times E_7$) is that obtained from the reduction of the $E_8 \times E_8$ heterotic string on $K3$ using the standard (non-standard) embedding(s) of the $K3$ instanton charge. So, in this search, we expect to obtain all solutions corresponding to these embeddings as well as the chains produced from these solutions by Higgsing. The solutions found are shown below.

\begin{enumerate}
\item $E_8 \times E_7$:
\begin{align}
&\textrm{(a) }10 (\mathbf{1},\mathbf{56}), \nn\\
&\textrm{(b) }\textstyle{\frac{3}{2}} (\mathbf{1},\mathbf{56}) + 4 (\mathbf{1},\mathbf{133}).
\end{align}
The first model on the list is the well-known model obtained from the reduction of the $E_8 \times E_8$ heterotic string on $K3$ using the standard embedding (24 units of instanton charge in one $E_8$).

\item $E_8 \times E_6$:
\begin{align}
& 18 (\mathbf{1},\mathbf{27}).
\end{align}
This solution, written in full as $9 (\mathbf{1},\mathbf{27}) + 9 (\mathbf{1},\overline{\mathbf{27}})$, may be obtained from the $E_8 \times E_7\textrm{(a)}$ models by Higgsing. The chain of Higgsing continues to further subgroups.

\item $E_8 \times F_4$:
\begin{align}
&\textrm{(a) } 17 (\mathbf{1},\mathbf{26}), \nn\\
&\textrm{(b) } 4 (\mathbf{1},\mathbf{52}) + 12 (\mathbf{1},\mathbf{26}).
\end{align}
\item $E_7 \times E_7$:
\begin{align}
&\textrm{(a) } \textstyle{\frac{n}{2}} (\mathbf{56},\mathbf{1}) + \textstyle{\frac{16-n}{2}} (\mathbf{1},\mathbf{56}); && n=0,\ldots,8, \nn\\
&\textrm{(b) } \textstyle{\frac{9}{2}} (\mathbf{56},\mathbf{1}) + n (\mathbf{133},\mathbf{1}) + 2 (\mathbf{1},\mathbf{56}); && n=0,1.
\end{align}
The first class of models on the list may be written in the more suggestive form $\textstyle{\frac{n_1-4}{2}} (\mathbf{56},\mathbf{1}) + \textstyle{\frac{n_2-4}{2}} (\mathbf{1},\mathbf{56})$, $n_1+n_2=24$ and they are recognized as the models constructed by reduction of the $E_8 \times E_8$ theory on $K3$ with $n_1$ and $n_2$ units of instanton charge embedded in the first and second $E_8$ respectively.

\item $E_7 \times E_6$:
\begin{align}
&\textrm{(a) } n (\mathbf{56},\mathbf{1}) + (14-2n) (\mathbf{1},\mathbf{27}); && n=0,\ldots,7, \nn\\
&\textrm{(b) } \textstyle{\frac{9}{2}} (\mathbf{56},\mathbf{1}) + 2 (\mathbf{1},\mathbf{27}), \nn\\
&\textrm{(c) } \textstyle{\frac{9}{2}} (\mathbf{56},\mathbf{1}) + (\mathbf{133},\mathbf{1}) + 2 (\mathbf{1},\mathbf{27}), \nn\\
&\textrm{(d) } 3(\mathbf{133},\mathbf{1}) + 2 (\mathbf{1},\mathbf{27}), \nn\\
&\textrm{(e) } \textstyle{\frac{n+2}{2}} (\mathbf{56},\mathbf{1}) + (5-n) (\mathbf{1},\mathbf{78}) + 2n (\mathbf{1},\mathbf{27}); && n=0,\ldots,2.
\end{align}
The first class of models are obtained from the $E_7 \times E_7 \textrm{(a)}$ models by Higgsing. The chain of Higgsing continues further on.

\item $E_7 \times F_4$:
\begin{align}
&\textrm{(a) } \textstyle{\frac{n}{2}} (\mathbf{56},\mathbf{1}) + (13-n) (\mathbf{1},\mathbf{26}); && n=0,\ldots,13, \nn\\
&\textrm{(b) } 2 (\mathbf{56},\mathbf{1}) + 6 (\mathbf{1},\mathbf{26}), \nn\\
&\textrm{(c) } \textstyle{\frac{9}{2}} (\mathbf{56},\mathbf{1}) + (\mathbf{1},\mathbf{26}), \nn\\
&\textrm{(d) } 2 (\mathbf{56},\mathbf{1}) + (\mathbf{1},\mathbf{52}) + 9 (\mathbf{1},\mathbf{26}), \nn\\
&\textrm{(e) } 2 (\mathbf{56},\mathbf{1}) + 3 (\mathbf{1},\mathbf{52}) + 6 (\mathbf{1},\mathbf{26}), \nn\\
&\textrm{(f) } 2 (\mathbf{56},\mathbf{1}) + 6 (\mathbf{1},\mathbf{52}), \nn\\
&\textrm{(g) } (\mathbf{133},\mathbf{1}) + \textstyle{\frac{9}{2}} (\mathbf{56},\mathbf{1}) + (\mathbf{1},\mathbf{26}), \nn\\
&\textrm{(h) } 3 (\mathbf{133},\mathbf{1}) + (\mathbf{1},\mathbf{26}), \nn\\
&\textrm{(i) } n (\mathbf{1},\mathbf{52}) + (9-n) (\mathbf{1},\mathbf{26}); && n=1,\ldots,6.
\end{align}

\item $E_6 \times E_6$:
\begin{align}
&\textrm{(a) } 2n (\mathbf{27},\mathbf{1}) + (12-2n) (\mathbf{1},\mathbf{27}); &&n=0,\ldots,6, \nn \\
&\textrm{(b) } 5 (\mathbf{78},\mathbf{1}); \nn\\
&\textrm{(c) } 2 (\mathbf{27},\mathbf{1}) + 4 (\mathbf{1},\mathbf{27}) + 3 (\mathbf{1},\mathbf{78}).
\end{align}

\item $F_4 \times F_4$:
\begin{align}
&\textrm{(a) } n (\mathbf{26},\mathbf{1}) + (10-n) (\mathbf{1},\mathbf{26}); && n=0,\ldots,5 \nn\\
&\textrm{(b) } n (\mathbf{26},\mathbf{1}) + (4-n) (\mathbf{1},\mathbf{52}) + (n+5) (\mathbf{1},\mathbf{26}); &&n=0,\ldots,4. \nn\\
&\textrm{(c) } (\mathbf{26},\mathbf{1}) + 6 (\mathbf{1},\mathbf{52}), \nn\\
&\textrm{(d) } (\mathbf{26},\mathbf{1}) + (\mathbf{1},\mathbf{52}) + 9 (\mathbf{1},\mathbf{26}).
\end{align}

\end{enumerate}

We finally proceed to the case where $\MG_1$ is an exceptional group while $\MG_2$ is classical. In this case,  $\MG_2$ does have fourth-order Casimirs and so we also have the extra condition (\ref{e-2-18}) for this factor. The models found are the following.

\begin{enumerate}
\item $E_8 \times \textrm{SU}(N)$:
\begin{align}
& 5 \leqslant N \leqslant 8: &&(112-12N) (\mathbf{1},\mathbf{N} ) + 14 \left(
\mathbf{1},\textstyle{\mathbf{\frac{N(N-1)}{2}}} \right).
\end{align}

\item $E_8 \times \textrm{SO}(N)$:
\begin{align}
& N=14: && 22 (\mathbf{1},\mathbf{14}) + 4 (\mathbf{1},\mathbf{64}).\\
& N=13: && 21 (\mathbf{1},\mathbf{13}) + 4 (\mathbf{1},\mathbf{64}).\\
& N=12: && \textrm{(a) } 20 (\mathbf{1},\mathbf{12}) + 8 (\mathbf{1},\mathbf{32}),\nn\\
                             &&&\textrm{(b) } 4 (\mathbf{1},\mathbf{66}) + 3 (\mathbf{1},\mathbf{12}) + \textstyle{\frac{15}{2}} (\mathbf{1},\mathbf{32}).\\
& N=11: && \textrm{(a) } 19 (\mathbf{1},\mathbf{11}) + 8 (\mathbf{1},\mathbf{32}),\nn\\
                             &&&\textrm{(b) } 4 (\mathbf{1},\mathbf{55}) + 6 (\mathbf{1},\mathbf{11}) + \textstyle{\frac{15}{2}} (\mathbf{1},\mathbf{32}).\\
& N=10: && \textrm{(a) }  18 (\mathbf{1},\mathbf{10}) + 16 (\mathbf{1},\mathbf{16}),\nn\\
                             &&&\textrm{(b) } 4 (\mathbf{1},\mathbf{45}) + 9 (\mathbf{1},\mathbf{10}) +  15 (\mathbf{1},\mathbf{16}).
\end{align}

\item $E_8 \times \textrm{Sp}(N)$:
\begin{align}
& N=4: &&16 (\mathbf{1},\mathbf{8} ) + 13 \left( \mathbf{1},\mathbf{27} \right).
\end{align}

\item $E_7 \times \textrm{SU}(N)$:
\begin{align}
&\!\!\!\!\!\!\!\!\!\!\!\!\!\!\!\! N=12: && 2 (\mathbf{56},\mathbf{1}) + 6
(\mathbf{1},\mathbf{66}).\\
&\!\!\!\!\!\!\!\!\!\!\!\!\!\!\!\! N=11: && 2 (\mathbf{56},\mathbf{1}) + 2 (\mathbf{1},\mathbf{11}) + 6 (\mathbf{1},\mathbf{55}).\\
&\!\!\!\!\!\!\!\!\!\!\!\!\!\!\!\! 5 \leqslant N \leqslant 10: && \textrm{(a) }  n (\mathbf{56},\mathbf{1}) +
\left[ 80-8N + 2n (N-8) \right] (\mathbf{1},\mathbf{N}) \nn\\
&&& + (10-2n) \left( \mathbf{1},\textstyle{\mathbf{\frac{N(N-1)}{2}}} \right); &&n=0,\ldots,5. \\
&\!\!\!\!\!\!\!\!\!\!\!\!\!\!\!\! N=5: && \textrm{(b) } \textstyle{\frac{n_1}{2}} (\mathbf{56},\mathbf{1}) + 4 n_1
(\mathbf{1},\mathbf{5}) + (7-n_1-2n_2) (\mathbf{1},\mathbf{24}) \nn\\
&&&+ 2n_2 (\mathbf{1},\mathbf{15}) +
(20-2n_1+2n_2) (\mathbf{1},\mathbf{10}); \nn\\
&&& n_1=0-7 , n_2 =0- \left\lfloor
\textstyle{\frac{7-n_1}{2}} \right\rfloor.
\end{align}

\item $E_7 \times \textrm{SO}(N)$:
\begin{align}
&\!\!\!\!\!\!\!\!\!\!\!\!\!\!\!\! 10 \leqslant N \leqslant 25: && \textrm{(a) } \textstyle{\frac{9}{2}} (\mathbf{56},\mathbf{1}) + (N-8) (\mathbf{1},\mathbf{N}).\\
&\!\!\!\!\!\!\!\!\!\!\!\!\!\!\!\! 10 \leqslant N \leqslant 19: && \textrm{(b) } 6 (\mathbf{56},\mathbf{1}) + (N-8) (\mathbf{1},\mathbf{N}).\\
&\!\!\!\!\!\!\!\!\!\!\!\!\!\!\!\! N=16: &&  \textrm{(c) } 2 (\mathbf{56},\mathbf{1}) + 16 (\mathbf{1},\mathbf{16}) + (\mathbf{1},\mathbf{128}).\\
&\!\!\!\!\!\!\!\!\!\!\!\!\!\!\!\! N=15: &&  \textrm{(c) } 2 (\mathbf{56},\mathbf{1}) + 15 (\mathbf{1},\mathbf{15}) + (\mathbf{1},\mathbf{128}).\\
&\!\!\!\!\!\!\!\!\!\!\!\!\!\!\!\! N=14: && \textrm{(c) }  (4-2n) (\mathbf{56},\mathbf{1}) + (4n+10) (\mathbf{1},\mathbf{14}) \nn\\
                              &&&+ (n+1)(\mathbf{1},\mathbf{64}); &&n=0,\ldots,2. \\
&\!\!\!\!\!\!\!\!\!\!\!\!\!\!\!\! 10 \leqslant N \leqslant 13: && \textrm{(c) } \textstyle{\frac{9}{2}} (\mathbf{56},\mathbf{1}) + (\mathbf{133},\mathbf{1}) + (N-8) (\mathbf{1},\mathbf{N}).\\
&\!\!\!\!\!\!\!\!\!\!\!\!\!\!\!\! N=13: && \textrm{(d) }  (6-n) (\mathbf{56},\mathbf{1}) + (2n+5) (\mathbf{1},\mathbf{13}) + \textstyle{\frac{n}{2}} (\mathbf{1},\mathbf{64}); &&n=0,\ldots,6. \\
&\!\!\!\!\!\!\!\!\!\!\!\!\!\!\!\! N=12: &&  \textrm{(d) } \textstyle{\frac{12-n}{2}} (\mathbf{56},\mathbf{1}) + (n+4) (\mathbf{1},\mathbf{12}) + \textstyle{\frac{n}{2}} (\mathbf{1},\mathbf{32}); &&n=0,\ldots,6. \\
&\!\!\!\!\!\!\!\!\!\!\!\!\!\!\!\! N=11: &&  \textrm{(d) } \textstyle{\frac{12-n}{2}} (\mathbf{56},\mathbf{1}) + (n+3) (\mathbf{1},\mathbf{11}) + \textstyle{\frac{n}{2}} (\mathbf{1},\mathbf{32}); &&n=0,\ldots,6, \nn\\
                              &&&\textrm{(e) } 3 (\mathbf{133},\mathbf{1}) + 3 (\mathbf{1},\mathbf{11}).\\
&\!\!\!\!\!\!\!\!\!\!\!\!\!\!\!\! N=10: && \textrm{(d) } \textstyle{\frac{12-n}{2}} (\mathbf{56},\mathbf{1}) + (n+2) (\mathbf{1},\mathbf{10}) + n (\mathbf{1},\mathbf{16}); &&n=0,\ldots,12, \nn\\
                              &&&\textrm{(e) } 3 (\mathbf{133},\mathbf{1}) + 2 (\mathbf{1},\mathbf{10}), \nn\\
                              &&&\textrm{(f) } 2 (\mathbf{56},\mathbf{1}) + (\mathbf{1},\mathbf{45}) + 9 (\mathbf{1},\mathbf{10}) + 9 (\mathbf{1},\mathbf{16}), \nn\\
                              &&&\textrm{(g) } \textstyle{\frac{n+2}{2}} (\mathbf{56},\mathbf{1}) + (5-n) (\mathbf{1},\mathbf{45}) \nn\\
                              &&&+ 2n (\mathbf{1},\mathbf{10}) + 8 (\mathbf{1},\mathbf{16}); &&n=0,\ldots,4.
\end{align}

\item $E_7 \times \textrm{Sp}(N)$:
\begin{align}
&\!\!\!\!\!\!\!\!\!\!\!\!\!\!\!\! 4 \leqslant N \leqslant 12: && \textrm{(a) }  2 (\mathbf{56},\mathbf{1}) + (24-2N) (\mathbf{1},\mathbf{2N}) \nn\\
                  &&&+ 2 (\mathbf{1},\mathbf{N(2N-1)-1}). \\
&\!\!\!\!\!\!\!\!\!\!\!\!\!\!\!\! N=6: && \textrm{(b) }  \textstyle{\frac{5}{2}} (\mathbf{56},\mathbf{1}) + 4 (\mathbf{1},\mathbf{12}) + 4 (\mathbf{1},\mathbf{65}) .\\
&\!\!\!\!\!\!\!\!\!\!\!\!\!\!\!\! N=5: && \textrm{(b) } \textstyle{\frac{9-n}{2}} (\mathbf{56},\mathbf{1}) + (18-2n) (\mathbf{1},\mathbf{10}) + n (\mathbf{1},\mathbf{44}); && n=0,\ldots,9.\\
&\!\!\!\!\!\!\!\!\!\!\!\!\!\!\!\! N=4: &&\textrm{(b) } \textstyle{\frac{9-n}{2}} (\mathbf{56},\mathbf{1}) + 16 (\mathbf{1},\mathbf{8}) + n (\mathbf{1},\mathbf{27}); && n=0,\ldots,9, \nn\\
                                      &&&\textrm{(c) } 2 (\mathbf{56},\mathbf{1}) + (\mathbf{1},\mathbf{36}) + 9 (\mathbf{1},\mathbf{27}) .
\end{align}

\item $E_6 \times \textrm{SU}(N)$:
\begin{align}
&\!\!\!\!\!\!\!\!\!\!\!\!\!\!\!\! N = 12: && 2 (\mathbf{27},\mathbf{1} ) + 6 (
\mathbf{1},\mathbf{66} ). \\
&\!\!\!\!\!\!\!\!\!\!\!\!\!\!\!\! N = 11: && \textrm{(a) } 2 (\mathbf{27},\mathbf{1} ) + 4 ( \mathbf{1},\mathbf{11} ) + 6 ( \mathbf{1},\mathbf{55} ), \nn\\
                          &&&\textrm{(b) } 4 (\mathbf{27},\mathbf{1} ) + 10 ( \mathbf{1},\mathbf{11} ) + 4 ( \mathbf{1},\mathbf{55} ). \\
&\!\!\!\!\!\!\!\!\!\!\!\!\!\!\!\! 5 \leqslant N \leqslant 10: && \textrm{(a) }  2n (\mathbf{27},\mathbf{1}) + \left[ 64-6N + 2n (N-8) \right] (\mathbf{1},\mathbf{N}) \nn\\
                                  &&& + (8-2n) \left( \mathbf{1},\textstyle{\mathbf{\frac{N(N-1)}{2}}} \right); &&n=0,\ldots,4. \\
&\!\!\!\!\!\!\!\!\!\!\!\!\!\!\!\! N = 5: && \textrm{(b) } (2n+1) (\mathbf{1},\mathbf{24} ) + 8 ( \mathbf{1},\mathbf{5} ) \nn\\
          &&& + (4-2n) ( \mathbf{1},\mathbf{15} ) + (20-2n) ( \mathbf{1},\mathbf{10} ); &&n=0,\ldots,2, \nn\\
                                         &&&\textrm{(c) } 2 (\mathbf{27},\mathbf{1} ) + (\mathbf{1},\mathbf{24} ) + 16 ( \mathbf{1},\mathbf{5} ) \nn\\
                                         &&&+ 2 ( \mathbf{1},\mathbf{15} ) + 14 ( \mathbf{1},\mathbf{10} ), \nn\\
                                         &&&\textrm{(d) } 2 (\mathbf{27},\mathbf{1} ) + 3(\mathbf{1},\mathbf{24} ) + 16 ( \mathbf{1},\mathbf{5} ) + 12 ( \mathbf{1},\mathbf{10} ), \nn\\
                                         &&&\textrm{(e) } 4 (\mathbf{27},\mathbf{1} ) + (\mathbf{1},\mathbf{24} ) + 24 ( \mathbf{1},\mathbf{5} ) + 8 ( \mathbf{1},\mathbf{10} ), \nn\\
                                         &&&\textrm{(f) } (\mathbf{78},\mathbf{1} ) + 8 (\mathbf{27},\mathbf{1} ) + 10 ( \mathbf{1},\mathbf{5} ).
\end{align}

\item $E_6 \times \textrm{SO}(N)$:
\begin{align}
&\!\!\!\!\!\!\!\!\!\!\!\!\!\!\!\! 10 \leqslant N \leqslant 20: && \textrm{(a) } 10 (\mathbf{27},\mathbf{1}) + (N-8) (\mathbf{1},\mathbf{N}).\\
&\!\!\!\!\!\!\!\!\!\!\!\!\!\!\!\! N=16: &&  \textrm{(b) } 2 (\mathbf{27},\mathbf{1}) + 16 (\mathbf{1},\mathbf{16}) + (\mathbf{1},\mathbf{128}).\\
&\!\!\!\!\!\!\!\!\!\!\!\!\!\!\!\! N=15: &&  \textrm{(b) } 2 (\mathbf{27},\mathbf{1}) + 15 (\mathbf{1},\mathbf{15}) + (\mathbf{1},\mathbf{128}).\\
&\!\!\!\!\!\!\!\!\!\!\!\!\!\!\!\! N=14: && \textrm{(b) }  (6-4n) (\mathbf{27},\mathbf{1}) + (4n+10) (\mathbf{1},\mathbf{14}) \nn\\
                                        &&&+ (n+1) (\mathbf{1},\mathbf{64}); &&n=0,1. \\
&\!\!\!\!\!\!\!\!\!\!\!\!\!\!\!\! N=13: && \textrm{(b) }  (10-2n) (\mathbf{27},\mathbf{1}) + (2n+5) (\mathbf{1},\mathbf{13}) + \textstyle{\frac{n}{2}} (\mathbf{1},\mathbf{64}); &&n=0,\ldots,5, \nn\\
&\!\!\!\!\!\!\!\!\!\!\!\!\!\!\!\! N=12: &&  \textrm{(b) } (10-n) (\mathbf{27},\mathbf{1}) + (2n+4) (\mathbf{1},\mathbf{12}) + n (\mathbf{1},\mathbf{32}); &&n=0,\ldots,5, \nn\\
                                        &&& \textrm{(c) } 2 (\mathbf{27},\mathbf{1}) + 9 (\mathbf{1},\mathbf{12}) + \textstyle{\frac{5}{2}} (\mathbf{1},\mathbf{32}), \nn\\
                                        &&& \textrm{(d) } 2 (\mathbf{27},\mathbf{1}) + (\mathbf{1},\mathbf{66}) + 9 (\mathbf{1},\mathbf{12}) + \textstyle{\frac{9}{2}} (\mathbf{1},\mathbf{32}) , \nn\\
                                        &&& \textrm{(e) } 2 (\mathbf{27},\mathbf{1}) + 3 (\mathbf{1},\mathbf{66}) + 4 (\mathbf{1},\mathbf{32}). \\
&\!\!\!\!\!\!\!\!\!\!\!\!\!\!\!\! N=11: &&  \textrm{(b) } (10-2n) (\mathbf{27},\mathbf{1}) + (2n+3) (\mathbf{1},\mathbf{11}) + n (\mathbf{1},\mathbf{32}); &&n=0,\ldots,5, \nn\\
                                        &&& \textrm{(c) } 2 (\mathbf{27},\mathbf{1}) + 8 (\mathbf{1},\mathbf{11}) + \textstyle{\frac{5}{2}} (\mathbf{1},\mathbf{32}), \nn\\
                                        &&& \textrm{(d) } 2 (\mathbf{27},\mathbf{1}) + (\mathbf{1},\mathbf{55}) + 9 (\mathbf{1},\mathbf{11}) + \textstyle{\frac{9}{2}} (\mathbf{1},\mathbf{32}) , \nn\\
                                        &&& \textrm{(e) } 2 (\mathbf{27},\mathbf{1}) + 3 (\mathbf{1},\mathbf{55}) + 2 (\mathbf{1},\mathbf{11}) + 4 (\mathbf{1},\mathbf{32}), \nn\\
                                        &&&\textrm{(f) } 3 (\mathbf{78},\mathbf{1}) + 4 (\mathbf{27},\mathbf{1})+ 3 (\mathbf{1},\mathbf{11}).\\
&\!\!\!\!\!\!\!\!\!\!\!\!\!\!\!\! N=10: && \textrm{(b) } (10-2n) (\mathbf{27},\mathbf{1}) + (n+2) (\mathbf{1},\mathbf{10}) + 2n (\mathbf{1},\mathbf{16}); &&n=0,\ldots,5, \nn\\
                                        &&&\textrm{(c) } 5 (\mathbf{1},\mathbf{45}) + 8 (\mathbf{1},\mathbf{16}), \nn\\
                                        &&&\textrm{(d) } 2 (\mathbf{27},\mathbf{1}) + (\mathbf{1},\mathbf{45}) + 9 (\mathbf{1},\mathbf{10}) + 9 (\mathbf{1},\mathbf{16}) , \nn\\
                                        &&&\textrm{(e) } 2 (\mathbf{27},\mathbf{1}) + 3 (\mathbf{1},\mathbf{45}) + 4 (\mathbf{1},\mathbf{10}) + 8 (\mathbf{1},\mathbf{16}), \nn\\
                                        &&&\textrm{(f) } (3-n) (\mathbf{78},\mathbf{1}) + (2n+4) (\mathbf{27},\mathbf{1}) \nn\\
                                        &&&+ (n+2) (\mathbf{1},\mathbf{10}) + n (\mathbf{1},\mathbf{16}); &&n=0,\ldots,2.
\end{align}

\item $E_6 \times \textrm{Sp}(N)$:
\begin{align}
&\!\!\!\!\!\!\!\!\!\!\!\!\!\!\!\!  4 \leqslant N \leqslant 12: && \textrm{(a) }  2 (\mathbf{27},\mathbf{1}) + (24-2N) (\mathbf{1},\mathbf{2N}) \nn\\
                  &&&+ 2 (\mathbf{1},\mathbf{N(2N-1)-1}). \\
&\!\!\!\!\!\!\!\!\!\!\!\!\!\!\!\!  N=6: && \textrm{(b) }  2 (\mathbf{27},\mathbf{1}) + 5 (\mathbf{1},\mathbf{65}) , \nn\\
       &&&\textrm{(c) }  4 (\mathbf{27},\mathbf{1}) + 8 (\mathbf{1},\mathbf{12}) + 3 (\mathbf{1},\mathbf{65}) . \\
&\!\!\!\!\!\!\!\!\!\!\!\!\!\!\!\!  N=5: &&  \textrm{(b) } (6-2n) (\mathbf{27},\mathbf{1}) + (16-4n) (\mathbf{1},\mathbf{10}) \nn\\
&&&+ (2n+1) (\mathbf{1},\mathbf{44}); && n=0,\ldots,3.\\
&\!\!\!\!\!\!\!\!\!\!\!\!\!\!\!\!  N=4: && \textrm{(b) } (6-2n) (\mathbf{27},\mathbf{1}) + 16 (\mathbf{1},\mathbf{8}) \nn\\
       &&&+ (2n+1) (\mathbf{1},\mathbf{27}); && n=0,\ldots,3, \nn\\
       &&&\textrm{(c) } 2 (\mathbf{27},\mathbf{1}) + (\mathbf{1},\mathbf{36}) + 9 (\mathbf{1},\mathbf{27}) .
\end{align}

\item $F_4 \times \textrm{SU}(N)$:
\begin{align}
&\!\!\!\!\!\!\!\!\!\!\!\!\!\!\!\! N = 12: && \textrm{(a) } (\mathbf{26},\mathbf{1} ) + 6 ( \mathbf{1},\mathbf{66} ), \nn\\
                          &&&\textrm{(b) } 3 (\mathbf{26},\mathbf{1} ) + 8 ( \mathbf{1},\mathbf{12} ) + 4 ( \mathbf{1},\mathbf{66} ). \\
&\!\!\!\!\!\!\!\!\!\!\!\!\!\!\!\! N = 11: && \textrm{(a) } (\mathbf{26},\mathbf{1} ) + 4 ( \mathbf{1},\mathbf{11} ) + 6 ( \mathbf{1},\mathbf{55} ), \nn\\
                          &&&\textrm{(b) } 3 (\mathbf{26},\mathbf{1} ) + 10 ( \mathbf{1},\mathbf{11} ) + 4 ( \mathbf{1},\mathbf{55} ), \nn\\
                          &&&\textrm{(c) } 5 (\mathbf{26},\mathbf{1} ) + 16 ( \mathbf{1},\mathbf{11} ) + 2 ( \mathbf{1},\mathbf{55} ). \\
&\!\!\!\!\!\!\!\!\!\!\!\!\!\!\!\! 5 \leqslant N \leqslant 10: && \textrm{(a) } (2n+1) (\mathbf{27},\mathbf{1}) + \left[ 48-4N + 2n (N-8) \right] (\mathbf{1},\mathbf{N}) \nn\\
                                  &&& + (6-2n) \left( \mathbf{1},\textstyle{\mathbf{\frac{N(N-1)}{2}}} \right); &&n=0,\ldots,3. \\
&\!\!\!\!\!\!\!\!\!\!\!\!\!\!\!\! N = 6: && \textrm{(b) } 3 (\mathbf{26},\mathbf{1} ) +  ( \mathbf{1},\mathbf{35} ) +  16 ( \mathbf{1},\mathbf{6} ) + 8 ( \mathbf{1},\mathbf{15} ). \\
&\!\!\!\!\!\!\!\!\!\!\!\!\!\!\!\! N = 5: && \textrm{(b) } 2n (\mathbf{1},\mathbf{24} ) + 12 ( \mathbf{1},\mathbf{5} ) \nn\\
          &&& + (4-2n) ( \mathbf{1},\mathbf{15} ) + (18-2n) ( \mathbf{1},\mathbf{10} ); &&n=0,\ldots,2, \nn\\
          &&&\textrm{(c) } (\mathbf{26},\mathbf{1} ) + (\mathbf{1},\mathbf{24} ) + 16 ( \mathbf{1},\mathbf{5} )\nn\\
          &&& + 2 ( \mathbf{1},\mathbf{15} ) + 14 ( \mathbf{1},\mathbf{10} ), \nn\\
          &&&\textrm{(d) } (4-n) (\mathbf{26},\mathbf{1} ) + n (\mathbf{1},\mathbf{24} ) \nn\\
          &&& + (28-4n) ( \mathbf{1},\mathbf{5} ) + (2n+6) ( \mathbf{1},\mathbf{10} ); &&n=0,\ldots,3, \nn\\
          &&&\textrm{(e) } 2 (\mathbf{26},\mathbf{1} ) + 20 ( \mathbf{1},\mathbf{5} ) + 2 ( \mathbf{1},\mathbf{15} ) + 12 ( \mathbf{1},\mathbf{10} ), \nn\\
          &&&\textrm{(f) } (\mathbf{52},\mathbf{1} ) + 8 (\mathbf{27},\mathbf{1} ) + 10 ( \mathbf{1},\mathbf{5} ).
\end{align}

\item $F_4 \times \textrm{SO}(N)$:
\begin{align}
&\!\!\!\!\!\!\!\!\!\!\!\!\!\!\!\! 10 \leqslant N \leqslant 20: && \textrm{(a) } 9 (\mathbf{26},\mathbf{1}) + (N-8) (\mathbf{1},\mathbf{N}).\\
&\!\!\!\!\!\!\!\!\!\!\!\!\!\!\!\! 10 \leqslant N \leqslant 25: && \textrm{(b) } 6 (\mathbf{26},\mathbf{1}) + (N-8) (\mathbf{1},\mathbf{N}).\\
&\!\!\!\!\!\!\!\!\!\!\!\!\!\!\!\! N=16:  && \textrm{(c) }  (\mathbf{26},\mathbf{1}) + 16 (\mathbf{1},\mathbf{16}) + (\mathbf{1},\mathbf{128}), \nn\\
                              &&&\textrm{(d) } (\mathbf{52},\mathbf{1}) + 9 (\mathbf{26},\mathbf{1}) + 8 (\mathbf{1},\mathbf{16}) . \\
&\!\!\!\!\!\!\!\!\!\!\!\!\!\!\!\! N=15: && \textrm{(c) } (\mathbf{26},\mathbf{1}) + 15 (\mathbf{1},\mathbf{15}) + (\mathbf{1},\mathbf{128}), \nn\\
                              &&&\textrm{(d) } (\mathbf{52},\mathbf{1}) + 9 (\mathbf{26},\mathbf{1}) + 7 (\mathbf{1},\mathbf{15}) . \\
&\!\!\!\!\!\!\!\!\!\!\!\!\!\!\!\! N=14: && \textrm{(c) }  (\mathbf{26},\mathbf{1}) + 14 (\mathbf{1},\mathbf{14}) + 2 (\mathbf{1},\mathbf{64}), \nn\\
                              &&&\textrm{(d) } 5 (\mathbf{26},\mathbf{1}) + 10 (\mathbf{1},\mathbf{14}) + (\mathbf{1},\mathbf{64}), \nn\\
                              &&&\textrm{(e) } (\mathbf{52},\mathbf{1}) + 9 (\mathbf{26},\mathbf{1}) + 6 (\mathbf{1},\mathbf{14}). \\
&\!\!\!\!\!\!\!\!\!\!\!\!\!\!\!\! 10 \leqslant N \leqslant 13: && \textrm{(c) } 9 (\mathbf{26},\mathbf{1}) + (\mathbf{52},\mathbf{1}) + (N-8) (\mathbf{1},\mathbf{N}).\\
&\!\!\!\!\!\!\!\!\!\!\!\!\!\!\!\! N=13: && \textrm{(d) } (7-2n) (\mathbf{26},\mathbf{1}) + (2n+7) (\mathbf{1},\mathbf{13}) \nn\\
                                        &&&+ \textstyle{\frac{n}{2}} (\mathbf{1},\mathbf{64}); &&n=0,\ldots,3. \\
&\!\!\!\!\!\!\!\!\!\!\!\!\!\!\!\! N=12: && \textrm{(d) } (8-n) (\mathbf{26},\mathbf{1}) + (n+5) (\mathbf{1},\mathbf{12}) \nn\\
                                        &&&+ \textstyle{\frac{n}{2}} (\mathbf{1},\mathbf{32}); &&n=0,\ldots,8, \nn\\
                              &&&\textrm{(e) } 13 (\mathbf{1},\mathbf{12}) + \textstyle{\frac{9}{2}} (\mathbf{1},\mathbf{32}), \nn\\
                              &&&\textrm{(f) } (\mathbf{26},\mathbf{1}) + 3 (\mathbf{1},\mathbf{66}) + 4 (\mathbf{1},\mathbf{32}),\nn\\
                              &&&\textrm{(g) } (\mathbf{26},\mathbf{1}) + 9 (\mathbf{1},\mathbf{12}) + \textstyle{\frac{5}{2}} (\mathbf{1},\mathbf{32}) ,\nn\\
                              &&&\textrm{(h) } (\mathbf{26},\mathbf{1}) + (\mathbf{1},\mathbf{66}) + 9 (\mathbf{1},\mathbf{12}) + \textstyle{\frac{9}{2}} (\mathbf{1},\mathbf{32}), \nn\\
                              &&&\textrm{(i) } 2 (\mathbf{26},\mathbf{1}) + 2 (\mathbf{1},\mathbf{66}) + 4 (\mathbf{1},\mathbf{12}) + 4 (\mathbf{1},\mathbf{32}),\nn\\
                              &&&\textrm{(j) } 2 (\mathbf{52},\mathbf{1}) + 7 (\mathbf{26},\mathbf{1}) + 5 (\mathbf{1},\mathbf{12}) + \textstyle{\frac{1}{2}} (\mathbf{1},\mathbf{32}),\nn\\
                              &&&\textrm{(k) } 3 (\mathbf{52},\mathbf{1}) + 6 (\mathbf{26},\mathbf{1}) + 4 (\mathbf{1},\mathbf{12}),\nn\\
                              &&&\textrm{(l) } 6 (\mathbf{52},\mathbf{1}) + 4 (\mathbf{1},\mathbf{12}).\\
&\!\!\!\!\!\!\!\!\!\!\!\!\!\!\!\! N=11: && \textrm{(d) } (8-n) (\mathbf{26},\mathbf{1}) + (n+4) (\mathbf{1},\mathbf{11}) + \textstyle{\frac{n}{2}} (\mathbf{1},\mathbf{32}); &&n=0,\ldots,8, \nn\\
                              &&&\textrm{(e) } 12 (\mathbf{1},\mathbf{11}) + \textstyle{\frac{9}{2}} (\mathbf{1},\mathbf{32}), \nn\\
                              &&&\textrm{(f) } (\mathbf{26},\mathbf{1}) + 8 (\mathbf{1},\mathbf{11}) + \textstyle{\frac{5}{2}} (\mathbf{1},\mathbf{32}),\nn\\
                              &&&\textrm{(g) } (\mathbf{26},\mathbf{1}) + (\mathbf{1},\mathbf{55}) + 9 (\mathbf{1},\mathbf{11}) + \textstyle{\frac{9}{2}} (\mathbf{1},\mathbf{32}) ,\nn\\
                              &&&\textrm{(h) } (n+1) (\mathbf{26},\mathbf{1}) + (3-n) (\mathbf{1},\mathbf{55})\nn\\
                              &&& + (3n+2) (\mathbf{1},\mathbf{11}) + 4 (\mathbf{1},\mathbf{32}); &&n=0,\ldots,2, \nn\\
                              &&&\textrm{(i) } 9 (\mathbf{26},\mathbf{1}) + 6 (\mathbf{1},\mathbf{11}) + \textstyle{\frac{3}{2}} (\mathbf{1},\mathbf{32}),\nn\\
                              &&&\textrm{(j) } (3-n) (\mathbf{52},\mathbf{1}) + (n+6) (\mathbf{26},\mathbf{1})\nn\\
                              &&& + (n+3) (\mathbf{1},\mathbf{11}) + \textstyle{\frac{n}{2}} (\mathbf{1},\mathbf{32}); &&n=0,\ldots,2, \nn\\
                              &&&\textrm{(k) } 6 (\mathbf{52},\mathbf{1}) + 3 (\mathbf{1},\mathbf{11}).\\
&\!\!\!\!\!\!\!\!\!\!\!\!\!\!\!\! N=10: && \textrm{(d) } (8-n) (\mathbf{26},\mathbf{1}) + (n+3) (\mathbf{1},\mathbf{11}) \nn\\
                                        &&&+ (n+1) (\mathbf{1},\mathbf{16}); &&n=0,\ldots,8, \nn\\
                              &&&\textrm{(e) } 4 (\mathbf{1},\mathbf{45}) + 2 (\mathbf{1},\mathbf{10}) + 8 (\mathbf{1},\mathbf{16}), \nn\\
                              &&&\textrm{(f) } (\mathbf{26},\mathbf{1}) + 7 (\mathbf{1},\mathbf{10}) + 5 (\mathbf{1},\mathbf{16}),\nn\\
                              &&&\textrm{(g) } (\mathbf{26},\mathbf{1}) + (\mathbf{1},\mathbf{45}) + 9 (\mathbf{1},\mathbf{10}) + 9 (\mathbf{1},\mathbf{16}) ,\nn\\
                              &&&\textrm{(h) } (n+1) (\mathbf{26},\mathbf{1}) + (3-n) (\mathbf{1},\mathbf{45}) \nn\\
                              &&& + (2n+4) (\mathbf{1},\mathbf{10}) + 8 (\mathbf{1},\mathbf{16}), \nn\\
                              &&&\textrm{(i) } 4 (\mathbf{26},\mathbf{1}) + 10 (\mathbf{1},\mathbf{10}) + 8 (\mathbf{1},\mathbf{16}),\nn\\
                              &&&\textrm{(j) } 9 (\mathbf{26},\mathbf{1}) + 5 (\mathbf{1},\mathbf{10}) + 3 (\mathbf{1},\mathbf{16}),\nn\\
                              &&&\textrm{(k) } (3-n) (\mathbf{52},\mathbf{1}) + (n+6) (\mathbf{26},\mathbf{1})\nn\\
                              &&& + (n+2) (\mathbf{1},\mathbf{10}) + n (\mathbf{1},\mathbf{16}); &&n=0,\ldots,2, \nn\\
                              &&&\textrm{(l) } 6 (\mathbf{52},\mathbf{1}) + 2 (\mathbf{1},\mathbf{10}).
\end{align}

\item $F_4 \times \textrm{Sp}(N)$:
\begin{align}
&\!\!\!\!\!\!\!\!\!\!\!\!\!\!\!\! 4 \leqslant N \leqslant 12: && \textrm{(a) } (\mathbf{26},\mathbf{1}) + (24-2N) (\mathbf{1},\mathbf{2N}) \nn\\
                                                              &&&+ 2 (\mathbf{1},\mathbf{N(2N-1)-1}). \\
&\!\!\!\!\!\!\!\!\!\!\!\!\!\!\!\! N=6: && \textrm{(b) }  2 (\mathbf{26},\mathbf{1}) + 4 (\mathbf{1},\mathbf{12}) + 4 (\mathbf{1},\mathbf{65}) , \nn\\
                                       &&&\textrm{(c) }  3 (\mathbf{26},\mathbf{1}) + 8 (\mathbf{1},\mathbf{12}) + 3 (\mathbf{1},\mathbf{65}) . \\
&\!\!\!\!\!\!\!\!\!\!\!\!\!\!\!\! N=5: && \textrm{(b) }  (6-n) (\mathbf{26},\mathbf{1}) + (18-2n) (\mathbf{1},\mathbf{10})+ n (\mathbf{1},\mathbf{44}); && n=0,\ldots,6.\\
&\!\!\!\!\!\!\!\!\!\!\!\!\!\!\!\! N=4: && \textrm{(b) } (6-n) (\mathbf{26},\mathbf{1}) + 16 (\mathbf{1},\mathbf{8}) + n (\mathbf{1},\mathbf{27}); && n=0,\ldots,3, \nn\\
                                       &&&\textrm{(c) } 3 (\mathbf{26},\mathbf{1}) + (\mathbf{1},\mathbf{36}) + 8 (\mathbf{1},\mathbf{27}) .
\end{align}
\end{enumerate}

Before concluding this section we note that, although we have not been able to make a thorough search for anomaly-free models when the gauge group is a product of two classical groups, a non-systematic search did not reveal any interesting models apart from those reported by Schwarz in \cite{Schwarz:1995zw} and some models related to them by Higgsing. For the sake of completeness, we list the basic models below.
\begin{enumerate}
\item $\textrm{SU}(N) \times \textrm{SU}(N)$. There exists the infinite class of models
\begin{align}
2 (\mathbf{N},\mathbf{N}).
\end{align}
\item $\textrm{SO}(N+8) \times \textrm{Sp}(N)$ with $0 \leqslant N \leqslant 24$. There exist the well-known small-instanton models
\begin{align}
\textstyle{\frac{1}{2}} (\mathbf{N+8},\mathbf{2N}) + \textstyle{\frac{24-N}{2}} (\mathbf{1},\mathbf{2N}) + (\mathbf{1},\mathbf{N(2N-1)-1}).
\end{align}
\item $\textrm{SO}(2N+8) \times \textrm{Sp}(N)$. There exists the infinite class of models
\begin{align}
(\mathbf{2N+8},\mathbf{2N}).
\end{align}
\end{enumerate}
The reader is referred to \cite{Schwarz:1995zw,Witten:1995gx} for more details on these models.

\section{Anomaly-free gauged supergravities}
\label{sec-4}

In this section, we continue our search, turning to the case of gauged supergravities where the R-symmetry group or a $\textrm{U}(1)_R$ subgroup thereof is gauged. The search for such models is of considerable interest due to the fact that these theories can spontaneously compactify on $\MBBR^4 \times \MBFS^2$ through a magnetic monopole background, leading to four-dimensional theories. In the case where the magnetic monopole is embedded in the R-symmetry group, stability \cite{Randjbar-Daemi:1983bw} of the compactification is ensured and the 4D theory is vectorlike. However, under certain conditions, it is also possible to embed the monopole in one of the other gauge group factors and obtain a chiral 4D spectrum. The aforementioned facts, as well as other interesting properties of the gauged models, provide enough motivation for looking for more consistent theories of this type. In fact, it is the search for anomaly-free gauged supergravities that motivated the work presented in this paper: given the fact that there is no known construction of such theories following from standard string/M-theory compactifications, the only way to identify consistent theories of this type is to directly solve the anomaly cancellation conditions.

The search for the gauged theories can be carried out in the same manner as before, this time including an extra $\textrm{Sp}(1)_R$ or $\textrm{U}(1)_R$ factor in the gauge group. So, the gauge group is now $\MG_s \times \MG_r$ and the new conditions that have to be satisfied are Eq. (\ref{e-2-36}) for the $\MG_r$ factor plus Eq. (\ref{e-2-37}) for the gauginos that transform nontrivially under both $\MG_s$ and $\MG_r$; using (\ref{e-2-11}) and (\ref{e-2-12}), we easily see that the first of these conditions is identically satisfied, leaving the second condition as the only non-trivial one. However, this last condition amounts to a set of strict equalities and, moreover, regarding the equalities involving $\MG_r$, the fact that the representations (or charges) of the fermions under this factor are fixed leaves little freedom for satisfying these constraints. So, one is led to expect that the gauged anomaly-free models will be very few.

The results of our search show that this is indeed the case. In the case of a gauge group of the type $\MG_1 \times \MG_r$, there is one equality constraint of the type (\ref{e-2-37}). In our search, we have not found any model solving the anomaly cancellation conditions. Passing to the case of a gauge group of the type $\MG_1 \times \MG_2 \times \MG_r$, there are three equality constraints of the type (\ref{e-2-37}) which are expected to seriously restrict the number of possible solutions. For the case where the whole $\textrm{Sp}(1)_R$ is gauged, we have found no solution. For the case where a $\textrm{U}(1)_R$ subgroup is gauged, we have found the following models.
\begin{enumerate}
\item $E_7 \times E_6 \times \textrm{U}(1)_R$ with the hypermultiplets transforming in
\begin{align}
\label{e-4-1}
\textstyle{\frac{1}{2}} (\mathbf{912},\mathbf{1}),
\end{align}
without singlet hypermultiplets. This is a well-known model, first found by Randjbar-Daemi, Salam, Sezgin and Strathdee in 1985. The important property of this model is that, besides the compactification with the monopole embedded in $\textrm{U}(1)_R$, it also admits a compactification with the monopole embedded in the ``hidden'' $E_6$, leading to an $\textrm{SO}(10) \times \textrm{SU}(2)_{KK}$ four-dimensional theory with chiral fermions. However, the demand for classical stability (no tachyonic modes) also fixes the monopole charge to its minimal value and restricts the number of families to two.
\item $E_7 \times G_2 \times \textrm{U}(1)_R$ with the hypermultiplets transforming in
\begin{align}
\label{e-4-2}
\textstyle{\frac{1}{2}} (\mathbf{56},\mathbf{14}),
\end{align}
again without singlets. This is a recently-found model, whose existence was reported in a recent paper \cite{Avramis:2005qt}. In that reference, the absence of both local and global anomalies was analytically demonstrated and various properties of the resulting supergravity theory were investigated. Unlike the previous model, this one does not admit any stable compactification with the monopole embedded in the $E_7 \times G_2$ factor, essentially because, in all possible embeddings of the monopole in this group, the absolute value $|q|$ of the $\textrm{U}(1)$ charge of the fermions takes more than one positive value. In particular, (i) the $E_7$ representation of the hypermultiplets is not the adjoint so that the decompositions of the type $E_7 \supset H \times \textrm{U}(1)$ lead to different values of $|q|$ for the $\textrm{U}(1)$--charged $E_7$ gauginos and hyperinos and (ii) although the $G_2$ gauginos and the hyperinos transform in the adjoint of $G_2$, this group does not have a decomposition of the type $G_2 \supset H \times \textrm{U}(1)$ with $H$ simple which would ensure that only one value of $|q|$ appears.

\item $F_4 \times \textrm{Sp}(9) \times \textrm{U}(1)_R$ with the hypermultiplets transforming in
\begin{align}
\label{e-4-3}
\textstyle{\frac{1}{2}} (\mathbf{52},\mathbf{18}),
\end{align}
again without singlets. This is a new model, first reported in this paper. Its basic phenomenological features, as far as $\MBBR^4 \times \MBFS^2$ compactifications are concerned, can be analyzed by following the guidelines of \cite{Randjbar-Daemi:1985wc,Avramis:2005qt}. Again, the model does not seem to admit a stable compactification with the monopole embedded in $F_4 \times \textrm{Sp}(9)$ because the $\textrm{Sp}(9)$ representation of the hypermultiplets is not the adjoint and $F_4$ is one of the few groups that has no decomposition of the type $F_4 \supset H \times \textrm{U}(1)$ with $H$ simple. The appearance of adjoint representations of groups with this particular property in this and the previous model is a curious coincidence.
\end{enumerate}

The structure of the models found is truly very interesting. In particular, they have the shared features that (i) the hypermultiplets transform in non-trivial representations (and, in the latter two cases, in product representations), (ii) there are no singlet hypermultiplets and (iii) the representations involve half-hypermultiplets. Moreover, as mentioned before, the cancellation of anomalies in these models is very delicate as can be verified by the explicit calculations of \cite{Randjbar-Daemi:1985wc} and \cite{Avramis:2005qt} for the former two. These facts might serve as indications that these gauged models are somehow related to critical string theory or M-theory by means of some mechanism. However, although some
progress has been made \cite{Cvetic:2003xr} regarding the archetypal Salam-Sezgin model, the origin of the models considered here remains mysterious up to date.
\\

\begin{small}
As mentioned in \S\ref{sec-2-4}, in the gauged case we have also allowed for an abelian gauge group factor $\MG_a$ that does not act on hypermultiplets. In the presence of such a factor, the gauge group includes ``drone'' $\textrm{U}(1)$'s under which all hypermultiplets and gauginos are singlets. Although this possibility leads to new anomaly-free models, the usual viewpoint is that turning on a large number of  $\textrm{U}(1)$'s so that the gravitational and R-symmetry anomalies are tuned to give a factorizable polynomial is quite \emph{ad hoc} and so these models are considered to be less important than the previous ones. Nevertheless, for reasons of completeness, we will list these models, for the case where the factor $\MG_s$ is simple and the number of $\textrm{U}(1)$'s is at most 50. The models found are the following.
\begin{enumerate}
\item $E_8 \times \textrm{U}(1)^3 \times \textrm{U}(1)_R$:
\begin{align}
\label{e-4-4}
2 \cdot \mathbf{248}.
\end{align}
\item $E_7 \times \textrm{U}(1)^{14} \times \textrm{Sp}(1)_R$:
\begin{align}
\label{e-4-5}
2 \cdot \mathbf{133} + 2 \cdot \mathbf{56}.
\end{align}
\item $E_7 \times \textrm{U}(1)^M \times \textrm{U}(1)_R$:
\begin{align}
& M=14: && 7 \cdot \mathbf{56}. \\
& M=18: && \mathbf{133} + \textstyle{\frac{9}{2}} \cdot \mathbf{56}. \\
& M=22: && 2 \cdot \mathbf{133} + 2 \cdot \mathbf{56}.
\end{align}
The $E_7 \times \textrm{U}(1)^{14} \times \textrm{U}(1)_R$ model has no singlets and is related to the $E_7 \times G_2 \times \textrm{U}(1)_R$ model of (\ref{e-4-2}) in the sense that the $G_2$ factor in the latter has been replaced by 14 $\textrm{U}(1)$'s. The existence of the $E_7 \times \textrm{U}(1)^{22} \times \textrm{U}(1)_R$ model was first pointed out in the footnote of \cite{Randjbar-Daemi:1985wc}.
\item $E_6 \times \textrm{U}(1)^{27} \times \textrm{Sp}(1)_R$:
\begin{align}
4 \cdot \mathbf{78}.
\end{align}
\item $E_6 \times \textrm{U}(1)^M \times \textrm{U}(1)_R$:
\begin{align}
& M=21: && 12 \cdot \mathbf{27}, \\
& M=29: && 2 \cdot \mathbf{78} + 6 \cdot \mathbf{27}, \\
& M=37: && 4 \cdot \mathbf{78}.
\end{align}
\item $\textrm{SU}(N) \times \textrm{U}(1)^M \times \textrm{U}(1)_R$:
\begin{align}
& N=8, M=42: && \mathbf{63} + 8 \cdot \mathbf{28}. \\
& N=7, M=45: && \mathbf{48} + 8 \cdot \mathbf{7} + 8 \cdot \mathbf{21}. \\
& N=6, M=46: && \mathbf{35} + 16 \cdot \mathbf{6} + 8 \cdot \mathbf{15}.
\end{align}
\item $\textrm{SU}(N) \times \textrm{U}(1)^M \times \textrm{U}(1)_R$:
\begin{align}
& 6 \leqslant N \leqslant 12, M = 8 + 12 N - N^2: && ( 48 - 4N ) \cdot \mathbf{N} + 6 \cdot \textstyle{\mathbf{\frac{N(N-1)}{2}}}. \\
&N=6,M=16: && 28 \cdot \mathbf{6} + 8 \cdot \mathbf{15}.
\end{align}
The first series of models have the same field content as the $SU(N)$(a) Poincar\'e theories found in \S\ref{sec-3-1} for $n=3$.
\item $\textrm{SO}(N) \times \textrm{U}(1)^M \times \textrm{Sp}(1)_R$:
\begin{align}
&N=10,M=12: && 12 \cdot \mathbf{11} + 10 \cdot \mathbf{16}.
\end{align}
\item $\textrm{SO}(N) \times \textrm{U}(1)^M \times \textrm{U}(1)_R$:
\begin{align}
&\!\!\!\!\!\!\!\!\!\!\!\!\!\!\!\! N=16,M=3: && 2 \cdot \mathbf{120} + \mathbf{128}. \\
&\!\!\!\!\!\!\!\!\!\!\!\!\!\!\!\! N=15,M=10: && 2 \cdot \mathbf{105} + \mathbf{15} + \mathbf{128}. \\
&\!\!\!\!\!\!\!\!\!\!\!\!\!\!\!\! N=14,M=16: && 2 \cdot \mathbf{91} + 2 \cdot \mathbf{14} + 2 \cdot \mathbf{64}. \\
&\!\!\!\!\!\!\!\!\!\!\!\!\!\!\!\! N=13,M=21: && 2 \cdot \mathbf{78} + 3 \cdot \mathbf{13} + 2 \cdot \mathbf{64}. \\
&\!\!\!\!\!\!\!\!\!\!\!\!\!\!\!\! N=12,M=17+4n: && n \cdot \mathbf{66} + (14 - 5n) \cdot \mathbf{12} + \textstyle{\frac{10-n}{2}} \cdot \mathbf{32}; && n=0,\ldots,2. \\
&\!\!\!\!\!\!\!\!\!\!\!\!\!\!\!\! N=11,M=20+4n: && n \cdot \mathbf{55} + (13-4n) \cdot \mathbf{11} + \textstyle{\frac{10-n}{2}} \cdot                        \mathbf{32}; && n=0,\ldots,3, \nn\\
&\!\!\!\!\!\!\!\!\!\!\!\!\!\!\!\! N=11,M=36: && 12 \cdot \mathbf{11} + \textstyle{\frac{9}{2}} \cdot \mathbf{32}. \\
&\!\!\!\!\!\!\!\!\!\!\!\!\!\!\!\! N=10,M=22+4n: && n \cdot \mathbf{45} + (12-3n) \cdot \mathbf{10} + (10-n) \cdot \mathbf{16}; && n=0,\ldots,4.
\end{align}
\item $\textrm{Sp}(N) \times \textrm{U}(1)^M \times \textrm{U}(1)_R$:
\begin{align}
&N=12,M=19: && 2 \cdot \mathbf{275}. \\
&N=11,M=42: && 2 \cdot \mathbf{22} + \mathbf{15} + \mathbf{230}. \\
&N=6,M=13: && 5 \cdot \mathbf{65}, \nn\\
&N=6,M=45: && 4 \cdot \mathbf{12} + 4 \cdot \mathbf{65}. \\
&N=5,M=24: && 8 \cdot \mathbf{10} + 5 \cdot \mathbf{44}.
\end{align}
\end{enumerate}
We see thus that allowing for the possibility of $\textrm{U}(1)$'s acting trivially on the hypermultiplets, we obtain many anomaly-free gauged models, some of which are extensions of the Poincar\'e models of \S\ref{sec-3-1}. Increasing the number of $\textrm{U}(1)$'s leads to numerous other models. However, as stressed above, these models are considered of limited interest.
\end{small}

\section{Discussion and outlook}
\label{sec-5}

In this paper, we have made a thorough search for anomaly-free $\MN=1$ supergravity theories in six dimensions, within the limits set by certain restrictions on the possible gauge groups and their representations. The search was made for both the Poincar\'e and gauged cases and all CPT-invariant hypermultiplet representations satisfying the anomaly cancellation conditions have been enumerated.

Our results are summarized as follows. In the Poincar\'e case, where there exist numerous solutions to the anomaly cancellation conditions, we have recovered most of the known models that have already been identified and constructed via various methods in the literature, plus a series of closely related models. We have also
found some models that have not been, to our knowledge, previously identified. Classifying these models and tracing their possible origin is outside the scope of the present paper.

In the gauged case, where the anomaly cancellation conditions are far more restrictive than in the Poincar\'e case, our search revealed the existence of just three models. The first is the well-known $E_7 \times E_6 \times \textrm{U}(1)_R$ model of \cite{Randjbar-Daemi:1985wc}, the second is an $E_7 \times G_2
\times \textrm{U}(1)_R$ model recently reported in \cite{Avramis:2005qt} and the third is an $F_4 \times
\textrm{Sp}(9) \times \textrm{U}(1)_R$ model discovered in this paper. All three models have an intriguing structure in the sense that the hypermultiplets transform in a single ``unusual'' representation of the gauge group with no singlets and, moreover, they satisfy the anomaly cancellation conditions in a ``miraculous'' manner. On the physical side, these models have very interesting properties, the most important one being the possibility of compactification to four dimensions through a monopole background with self-tuning of the cosmological constant. These compactifications reveal, however, some phenomenological problems with these theories, for example the fact that the demand for stability of these compactifications leads either to too few or too many families. Allowing for the presence of extra ``drone'' $\textrm{U}(1)$ factors, we have identified many more anomaly-free gauged models. However, the presence of the extra $\textrm{U}(1)$'s renders these models
less elegant than those described earlier.

The search presented in this paper can be extended towards several directions, the main focus being on finding new consistent gauged theories. For instance, one may consider gauge groups that contain three or more simple factors. Also, one may consider theories with more than one tensor multiplet, where there exists the generalized Green-Schwarz mechanism that allows anomaly freedom under weaker constraints. One could finally consider adding extra $\textrm{U}(1)$ factors that act non-trivially on the hypermultiplets but, unless there is a physical principle that determines the $\textrm{U}(1)$ charges in some way, this is a very complicated task. We hope that the work presented here will initiate some progress along these lines.

\acknowledgments

The authors wish to thank S. Randjbar-Daemi for enlightening discussions during the course of this work. This work is supported by the EPEAEK programme ``Heraclitus'' and the NTUA programme ``Protagoras''.

\appendix

\section{Anomaly polynomials}
\label{appa}

The anomaly polynomials used in this paper are normalized as follows
\begin{eqnarray}
\label{e-a-1}
I^{1/2}_{8} (R) &=& \frac{1}{360} \tr R^4 + \frac{1}{288} (\tr R^2)^2, \nn\\
I^{1/2}_{8} (F) &=& \frac{2}{3} \tr F^4, \nn\\
I^{1/2}_{8} (F_A,F_B) &=& 4 \tr F_A^2 \tr F_B^2, \nn\\
I^{1/2}_{8} (F,R) &=& - \frac{1}{6} \tr R^2 \tr F^2, \nn\\
I^{3/2}_{8} (R) &=& \frac{49}{72} \tr R^4 - \frac{43}{288} (\tr R^2)^2 ,\nn\\
I^{3/2}_{8} (F) &=& \frac{10}{3} \tr F^4, \nn\\
I^{3/2}_{8} (F,R) &=& \frac{19}{6} \tr R^2 \tr F^2, \nn\\
I^A_{8} (R) &=& - \frac{7}{90} \tr R^4 + \frac{1}{36} (\tr R^2)^2.
\end{eqnarray}
Here, the superscripts $1/2$, $3/2$ and $A$ refer to a spin $1/2$ fermion, a spin $3/2$ fermion and a 2--form
potential respectively. The above anomaly polynomials correspond to Weyl spinors of positive chirality and 2--form potentials with self-dual field strengths. For a spinor subject to a Majorana-type condition, one needs to include a factor of $\frac{1}{2}$, while for a negative-chirality spinor or an anti-self-dual field strength the sign of the anomaly is reversed.

\section{Group-theoretical coefficients}
\label{appb}

Here we give, for reference purposes, the formulas for the group-theoretical coefficients $a$, $b$ and $c$ appearing in the discussion of Section \ref{sec-2}. We consider a simple group $\MG$ and we let $\MR$, $\MF$ and $\MA$ be a generic representation, the fundamental and the adjoint respectively. The $n$--th index $\ell_n(\MR)$ of $\MR$ is defined in terms of the symmetrized trace of the product of $n$ generators. In particular the second and fourth indices are determined by
\begin{equation}
\label{e-b-1}
\Str_{\MR} T^a T^b = \ell_2(\MR) d^{ab},
\end{equation}
and
\begin{equation}
\label{e-b-2}
\Str_{\MR} T^a T^b T^c T^d = \ell_4( \MR ) d^{abcd}  + \frac{3}{2 + \dim \MA } \ell_2( \MR )^2 \left[ \frac{\dim \MA}{\dim \MR} - \frac{1}{6} \frac{\ell_2( \MA )}{\ell_2 ( \MR )} \right] d^{(ab}d^{cd)}.
\end{equation}
where $d^{a_1 \ldots a_n}$ are the invariant symmetric tensors of $\MG$ subject to the orthogonality conditions $d^{a_1 \ldots a_m} d^{a_1 \ldots a_m \ldots a_n} = 0$ for $m<n$; their normalization is determined by fixing the values of $\ell_n ( \MF )$. The normalization of second-order indices is irrelevant for our purposes while the normalization of fourth-order indices can be fixed by setting $\ell_4(\MF) = 1$ for all groups.

\begin{table}[!t]
\begin{center}
\begin{tabular}{|c||c|c|c|}
\hline
Group & Irrep $\MR$ & $b_{\MR}$ & $c_{\MR}$ \\
\hline
\hline
$E_8$ & $\mathbf{248}$ & $1/100$ & $1$ \\
\hline
      & $\mathbf{56}$ & $1/24$ & $1$ \\
\cline{2-4}
$E_7$ & $\mathbf{133}$ & $1/6$ & $3$ \\
\cline{2-4}
      & $\mathbf{912}$ & $31/12$ & $30$ \\
\hline
      & $\mathbf{27}$ & $1/12$ & $1$ \\
\cline{2-4}
      & $\mathbf{78}$ & $1/2$ & $4$ \\
\cline{2-4}
$E_6$ & $\mathbf{351}$ & $55/12$ & $25$ \\
\cline{2-4}
      & $\mathbf{351'}$ & $35/6$ & $28$ \\
\cline{2-4}
      & $\mathbf{650}$ & $10$ & $50$ \\
\hline
      & $\mathbf{26}$ & $1/12$ & $1$ \\
\cline{2-4}
$F_4$ & $\mathbf{52}$ & $5/12$ & $3$ \\
\cline{2-4}
      & $\mathbf{273}$ & $49/12$ & $21$ \\
\cline{2-4}
      & $\mathbf{324}$ & $23/4$ & $27$ \\
\hline
      & $\mathbf{7}$ & $1/4$ & $1$ \\
\cline{2-4}
$G_2$ & $\mathbf{14}$ & $5/2$ & $4$ \\
\cline{2-4}
      & $\mathbf{27}$ & $27/4$ & $9$ \\
\cline{2-4}
      & $\mathbf{64}$ & $38$ & $32$ \\
\hline
        & $\mathbf{3}$ & $1/2$ & $1$ \\
\cline{2-4}
$\textrm{SU}(3)$ & $\mathbf{8}$ & $9$ & $6$ \\
\cline{2-4}
        & $\mathbf{6}$ & $17/2$ & $5$ \\
\hline
$\textrm{SU}(2)$ & $\mathbf{2}$ & $1/2$ & $1$ \\
\cline{2-4}
        & $\mathbf{3}$ & $8$ & $4$ \\
\hline
\end{tabular}
\end{center}
\caption{The coefficients $b$ and $c$ for groups with no fourth-order invariants.}
\label{t-2}
\end{table}

To compute the $c$--coefficients, we consider an algebra element $X=X^a T^a$ and we use (\ref{e-b-1}) for the representations $\MR$ and $\MF$ to find
\begin{equation}
\label{e-b-3}
\tr_{\MR} X^2 = \ell_2(\MR) (X^a)^2 ,\qquad \tr_{\MF} X^2 = \ell_2(\MF) (X^a)^2
\end{equation}
where we use the notation $(X^a)^n \equiv d^{a_1 \ldots a_n} X^{a_1} \ldots X^{a_n}$. So, we have
\begin{equation}
\label{e-b-4}
c_{\MR} =  \frac{\ell_2 ( \MR )}{\ell_2 ( \MF )} .
\end{equation}
To compute the $a$-- and $b$--coefficients, we first consider the case where $\MR$ has no fourth-order Casimirs so that $\ell_4(\MR) = 0$. Then Eq. (\ref{e-b-2}) leads to
\begin{equation}
\label{e-b-5}
\tr_{\MR} X^4 = \frac{3}{2 + \dim \MA } \ell_2( \MR )^2 \left[ \frac{\dim \MA}{\dim \MR} - \frac{1}{6} \frac{\ell_2( \MA )}{\ell_2 ( \MR )} \right] ((X^a)^2)^2 ,
\end{equation}
and so, $a_{\MR}=0$ and
\begin{equation}
\label{e-b-6}
b_{\MR} = \frac{3}{2 + \dim \MA } \frac{\ell_2( \MR )^2}{\ell_2( \MF )^2} \left[ \frac{\dim \MA}{\dim \MR} - \frac{1}{6} \frac{\ell_2( \MA )}{\ell_2 ( \MR )} \right] .
\end{equation}
We next consider the case where $\MR$ possesses fourth-order Casimirs. Then, using (\ref{e-b-2}) for the representations $\MR$ and $\MF$, we find
\begin{equation}
\label{e-b-7}
\tr_{\MR} X^4 = \ell_4( \MR ) (X^a)^4  + \frac{3}{2 + \dim \MA } \ell_2( \MR )^2 \left[ \frac{\dim \MA}{\dim \MR} - \frac{1}{6} \frac{\ell_2( \MA )}{\ell_2 ( \MR )} \right] ((X^a)^2)^2 ,
\end{equation}
and
\begin{equation}
\label{e-b-8}
\tr_{\MF} X^4 = (X^a)^4  + \frac{3}{2 + \dim \MA } \ell_2( \MF )^2 \left[ \frac{\dim \MA}{\dim \MF} - \frac{1}{6} \frac{\ell_2( \MA )}{\ell_2 ( \MF )} \right] ((X^a)^2)^2 ,\qquad
\end{equation}
Solving (\ref{e-b-8}) for $(X^a)^4$, substituting in (\ref{e-b-7}) and using the second of (\ref{e-b-3}), we find
\begin{equation}
\label{e-b-9}
a_{\MR} = \ell_4( \MR ) ,
\end{equation}
\begin{equation}
\label{e-b-10}
b_{\MR} \!=\! \frac{3}{2 + \dim \MA } \left\{ \frac{\ell_2( \MR )^2}{\ell_2( \MF )^2} \left[ \frac{\dim \MA}{\dim \MR} - \frac{1}{6} \frac{\ell_2( \MA )}{\ell_2 ( \MR )} \right] \!-\! \ell_4(\MR) \left[ \frac{\dim \MA}{\dim \MF} - \frac{1}{6} \frac{\ell_2( \MA )}{\ell_2 ( \MF )} \right] \right\}.
\end{equation}
\begin{table}[!t]
\begin{center}
\begin{tabular}{|c||c|c|c|c|}
\hline
Group & Irrep $\MR$ & $a_{\MR}$ &$b_{\MR}$ & $c_{\MR}$\\
\hline
\hline
& $\mathbf{N}$ & $1$ & $0$ & $1$\\
\cline{2-5}
$\textrm{SU}(N)$ & $\mathbf{N^2-1}$ & $2N$ & $6$ & $2N$\\
\cline{2-5}
& $\textstyle{\mathbf{\frac{N(N+1)}{2}}}$ & $N+8$ & $3$ & $N+2$\\
\cline{2-5}
& $\textstyle{\mathbf{\frac{N(N-1)}{2}}}$ & $N-8$ & $3$ & $N-2$\\
\hline
\hline
& $\mathbf{N}$ & $1$ & $0$ & $1$ \\
\cline{2-5}
$\textrm{SO}(N)$ & $\textstyle{\mathbf{\frac{N(N-1)}{2}}}$ & $N-8$ & $3$ & $N-2$\\
\cline{2-5}
& $\mathbf{2^{\lfloor \frac{N+1}{2} \rfloor - 1}}$ & $- 2^{\lfloor \frac{N+1}{2} \rfloor -5} $ & $3 \cdot 2^{\lfloor \frac{N+1}{2} \rfloor -7} $ & $2^{\lfloor \frac{N+1}{2} \rfloor - 4} $\\
\hline
\hline
& $\mathbf{2N}$ & $1$ & $0$ & $1$\\
\cline{2-5}
$\textrm{Sp}(N)$ & $\mathbf{N(2N+1)}$ & $2N+8$ & $3$ & $2N+2$\\
\cline{2-5}
& $\mathbf{N(2N-1)-1}$ & $2N-8$ & $3$ & $2N-2$\\
\hline
\end{tabular}
\end{center}
\caption{The coefficients $a$, $b$ and $c$ for for groups with fourth-order invariants.}
\label{t-3}
\end{table}

From these expressions, one may determine all the group-theoretical coefficients of interest using the values of the indices $\ell_2(\MR)$ and $\ell_4(\MR)$ which are tabulated e.g. in \cite{Slansky:1981yr}, \cite{McKay}, \cite{vanRitbergen:1998pn}. The values of $b_{\MR}$ and $c_{\MR}$ for groups with no fourth-order Casimirs are listed on Table \ref{t-2}. The values of $a_{\MR}$, $b_{\MR}$ and $c_{\MR}$ for groups having fourth-order Casimirs are listed on Table \ref{t-3}.

}

\end{document}